%% file: oneLoopColor.tex
\documentclass[12pt,a4paper]{article}
\pdfoutput=1
\usepackage{jheppub,psfrag,slashed,cancel,lscape,caption,array,graphicx,subcaption}
\usepackage[utf8]{inputenc}
\usepackage{amsmath}
\usepackage{amssymb}
\usepackage{mathtools}
\usepackage{mathrsfs}
\usepackage{multirow}
\usepackage{booktabs} 
\usepackage{xcolor}
\usepackage{bm}
\usepackage{mathtools}
\usepackage{tikz,fp}
\usetikzlibrary{decorations.pathmorphing,positioning,snakes,calc,fadings}
\usetikzlibrary{decorations.markings,decorations.pathreplacing}

\tikzset{
  boson/.style={
    decoration={snake, segment length=2mm, amplitude=0.5mm},
    decorate,
  },
  fermion/.style={
    postaction={decorate},
    decoration={markings,mark=at position .6 with {\arrow[scale=1.5]{>}}}
  },
  gluon/.style={
    decorate,
    decoration={coil,amplitude=2.4pt,segment length=2.7pt}
  },
  scalar/.style={
    dashed
  },
  directedScalar/.style={
    dashed,
    postaction={decorate},
    decoration={markings,mark=at position 0.5 with {\arrow[scale=1.5]{>}}}
  },
  directedBoson/.style={
    fixed point arithmetic,
    decoration={snake, segment length=2mm, amplitude=0.5mm},
    decorate,
    postaction={
      decoration={
        markings,
        mark=at position 0.5 with {\arrow[scale=1.5]{>}}
      },
      decorate
    }
  },
  middleArrow/.style={
    draw=white,
    decoration={markings,mark=at position 0.5 with{\arrow[scale=1.5,black]{>}}},
    decorate
  }
}

\newcommand{\cC}[1]{\node[circle, fill=white, draw=black, inner sep=1.3] at #1 {}}

\DeclareFontFamily{OT1}{pzc}{}
\DeclareFontShape{OT1}{pzc}{m}{it}{<-> s * [1.350] pzcmi7t}{}
\DeclareMathAlphabet{\mathpzc}{OT1}{pzc}{m}{it}

\definecolor{mygreen}{rgb}{0,0.4,0}

\def\cT{\mathcal{T}}

\newcommand{\ol}{\overline}
\newcommand{\ul}{\underline}
\newcommand{\tR}{\text{R}}
\newcommand{\tL}{\text{L}}
\DeclareMathOperator{\Tr}{Tr}
\DeclareMathSymbol{\mlq}{\mathord}{operators}{``}
\DeclareMathSymbol{\mrq}{\mathord}{operators}{`'}

\DeclareFontFamily{U}{mathx}{\hyphenchar\font45}
\DeclareFontShape{U}{mathx}{m}{n}{<-> mathx10}{}
\DeclareSymbolFont{mathx}{U}{mathx}{m}{n}
\DeclareMathAccent{\widebar}{0}{mathx}{"73}

\preprint{UUITP-48/17}

\title{Cyclic Mario Worlds -- Color-Decomposition for One-Loop QCD}

\author[a]{Gregor K\"{a}lin}
\affiliation[a]{Department of Physics and Astronomy, Uppsala University,\\ Box 516, SE-751 20 Uppsala, Sweden}
\emailAdd{gregor.kaelin@physics.uu.se}

\abstract{
We present a new color decomposition for QCD amplitudes at one-loop level as a generalization of the Del Duca-Dixon-Maltoni and Johansson-Ochirov decomposition at tree level. Starting from a minimal basis of planar primitive amplitudes we write down a color decomposition that is free of linear dependencies among appearing primitive amplitudes or color factors. The conjectured decomposition applies to any number of quark flavors and is independent of the choice of gauge group and matter representation. The results also hold for higher-dimensional or supersymmetric extensions of QCD. We provide expressions for any number of external quark-antiquark pairs and gluons.
\begin{figure}[b]
  \begin{center}
    \includegraphics{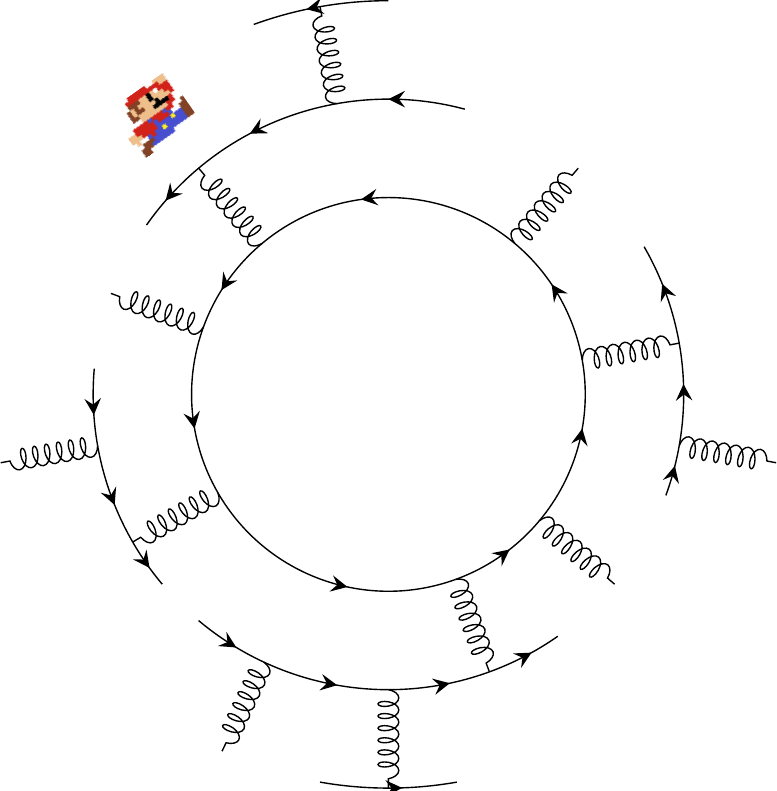} 
  \end{center}
\end{figure}
}


\begin{document}
\maketitle
\flushbottom

\section{Introduction}

In the past two decades, in anticipation of LHC, we have seen some spectacular progress in the computation of next-to-leading order (NLO) processes with higher multiplicities \cite{Bern:1990cu, Britto:2004nc, Forde:2007mi, Ossola:2006us, Anastasiou:2006jv, Anastasiou:2006gt, Giele:2008ve, Giele:2008bc, Ellis:2008ir, Ellis:2008qc, Assadsolimani:2009cz, Becker:2010ng, Becker:2012aqa, Ellis:2011cr, Bern:2013gka, Bonciani:2015eua}. New methods, like the (generalized) unitarity method and applications thereof~\cite{Bern:1994zx, Bern:1994cg, Bern:1996je, Ellis:2007br, Berger:2008sj, Ossola:2007ax, Mastrolia:2008jb, Mastrolia:2010nb, Badger:2010nx, Hirschi:2011pa}, recursive constructions~\cite{Berends:1987me, Britto:2004ap, Britto:2005fq, Mafra:2010jq}, MHV rules~\cite{Cachazo:2004kj} or integration technology~\cite{vanOldenborgh:1989wn, Gehrmann:1999as, Anastasiou:2002yz, Anastasiou:2003yy, Anastasiou:2005cb, Ellis:2007qk, Smirnov:2012gma, Duhr:2011zq, Anastasiou:2013srw, Caola:2014lpa, Henn:2014loa, Henn:2014qga, Zhang:2016kfo} have significantly improved the accessible region of NLO computations.

A standard technique in modern methods for computing tree level scattering amplitudes in quantum chromodynamics (QCD) is the decomposition of the amplitudes into purely kinematic primitive amplitudes and purely color-dependent objects (color factors). For tree-level processes the primitive, color-ordered amplitudes are often computed using recursion in terms of the number of legs~\cite{Berends:1987me, Cachazo:2004kj, Britto:2004ap, Britto:2005fq}. The decomposition into color and kinematic parts is not unique -- and it is not obvious how to find the most useful and compact formulae.

The standard $\text{SU}(N)$ trace-based color decomposition~\cite{Mangano:1988kk, Berends:1987cv, Kosower:1987ic, Mangano:1987xk, Bern:1990ux} at tree level does not take advantage of all linear dependencies of primitive amplitudes and color factors, as the sum goes over an overcomplete set of linear dependent primitive amplitudes and color factors. The same holds for amplitudes with multi-quark external states~\cite{Cvitanovic:1980bu, Maltoni:2002mq, Weinzierl:2005dd, Mangano:1990by, Reuschle:2013qna}. The linear relations arise from the color algebra of gauge theory~\cite{Kleiss:1988ne} and observed linear relations between primitive amplitudes: the Kleiss-Kuijf (KK)~\cite{Kleiss:1988ne} and the Bern-Carrasco-Johansson (BCJ) relations~\cite{Bern:2008qj}. 

These relations have been utilized to find more compact color decompositions, first for the purely gluonic case by Del-Duca-Dixon-Maltoni (DDM)~\cite{DelDuca:1999iql,DelDuca:1999rs} and later the generalization to any number of external quark-antiquark pairs by Johansson and Ochirov (JO)~\cite{Johansson:2015oia}, which was proven by Melia~\cite{Melia:2015ika} shortly thereafter.

At one-loop level color decompositions have been worked out for specific cases~\cite{Bern:1994fz, DelDuca:1999rs, Ellis:2011cr, Ita:2011ar, Badger:2012pg}, in a general trace-based setup~\cite{Reuschle:2013qna, Schuster:2013aya} including an overcomplete set of primitive amplitudes and recently Ochirov and Page presented a general method to obtain the full color dependence at loop level~\cite{Ochirov:2016ewn}. In this paper we propose a compact color decomposition that eliminates linear dependencies of one-loop QCD amplitudes to general multiplicity and number of external quark-antiquark pairs. The structure of the planar diagrams contributing to a primitive amplitude and its corresponding color factor in the color decomposition have a similar form as in the tree level case -- the `Mario world' diagrams~\cite{Johansson:2015oia, Melia:2015ika}. For one-loop diagrams internal lines carrying loop momentum form a cyclic `ground level' for the Mario world and the external partons build up the `higher levels' -- mathematically the different parts in the tensor product of the Lie algebra. A basis of one-loop primitive amplitudes is given by all such Mario world structures, for both cases where either a closed quark or a gluonic loop is present in the diagrams contributing to a primitive amplitude. For $n$ external partons, whereof $k$ are quark-antiquark pairs, the size of the sets of primitive amplitudes considered in the color decomposition is $(n-1)!/k!$ for the case of a closed quark loop -- or $(n-2)!(n-k-1)/k!$ if the gauge group has traceless Lie algebra generators -- and $2^{2k-1}(n-1)!k!/(2k)!$ if at least one gluon carries loop momentum.

We start by setting up the notation and review the tree-level color decomposition in section~\ref{sec:review}. Section~\ref{sec:basis} defines a basis of primitive one-loop amplitudes for the two cases where either a closed quark loop or a mixed/gluonic loop is present in the amplitude. In what follows we present a color decomposition to all multiplicities in section~\ref{sec:decomp} together with a pedagogical 5-point example in section~\ref{sec:ex}. We discuss the results and give an outlook in section~\ref{sec:conclusions}.

\section{Tree level review and prerequisites}\label{sec:review}

We review the work by Johansson and Ochirov (JO)~\cite{Johansson:2015oia} about a new color decomposition for massless QCD at tree level and introduce the necessary notation.

Del Duca, Dixon and Maltoni (DDM)~\cite{DelDuca:1999iql,DelDuca:1999rs} presented a color decomposition for gluonic amplitudes that removes the redundancy of Kleiss-Kuijf (KK) relations~\cite{Kleiss:1988ne} -- present in the familiar trace-based $\text{SU}(N_c)$ color decomposition~\cite{Mangano:1988kk, Berends:1987cv, Mangano:1987xk, Bern:1990ux}. The decomposition is written in terms of gauge group structure constants~$\tilde{f}^{abc}$ and primitive (color-ordered) amplitudes as
\begin{equation}
  \mathcal{A}^{(0)}_n = g^{n-2}\smashoperator{\sum_{\sigma\in S_{n-2}(\{3,\dots,n\})}}\tilde{f}^{a_2 a_{\sigma(3)} b_1} \tilde{f}^{b_1 a_{\sigma(4)}b_2}\cdots \tilde{f}^{b_{n-3}a_{\sigma(n)}a_1}A(1,2,\sigma),
\end{equation}
where we define the structure constants in terms of the gauge group generators~$T^a$
\begin{align}
   &\tilde{f}^{abc} = \Tr([T^a,T^b]T^c),& &\Tr(T^aT^b) = \delta^{ab}.
\end{align}
The primitive amplitudes~$A(1,2,\sigma)$ are exactly the color-ordered amplitudes appearing in the trace-based color decomposition -- and can be directly computed from planar diagrams with the given external ordering using color-ordered Feynman rules~\cite{Dixon:1996wi}. This decomposition is valid for any choice of gauge group since it only relies on defining properties of the Lie algebra.

\subsection{A basis for primitive QCD tree amplitudes}
A generalization of the above color decomposition that includes any number of external distinguishable quark-antiquark pairs builds on a basis of primitive tree amplitudes presented by Melia~\cite{Melia:2013bta, Melia:2014oza}. As in the purely gluonic case primitive amplitudes are defined as the sum over all planar diagrams with the given cyclic ordering of externals using color-ordered Feynman rules. For example we can diagrammatically represent a primitive amplitude as
\begin{equation}\label{eqn:ex_tree_prim}
  A^{(0)}(\ul{1},\ol{1},\ol{2},3,\ul{2},\ol{4},\ol{5},6,\ul{5},\ul{4}) \, = \,
  \begin{tikzpicture}
    [>=stealth, baseline=-0.5, scale=0.7]
    \draw[dashed] (0:2) arc (0:360:2);
    \draw[fermion] (-36:2.5) node[right] {$\ol{1}$} .. controls (-36:1) and (0:1) .. (0:2.5) node[right] {$\ul{1}$};
    \draw[fermion] (180:2.5) node[left] {$\ol{4}$} .. controls (180:1) and (36:1) .. (36:2.5) node[right] {$\ul{4}$};
    \draw[fermion] (144:2.5) node[left] {$\ol{5}$} .. controls (144:1) and (72:1) .. (72:2.5) node[above] {$\ul{5}$};
    \draw[gluon] (108:2.5) node[above] {$6$} -- (108:1.5);
    \draw[fermion] (288:2.5) node[below] {$\ol{2}$} .. controls (288:1) and (216:1) .. (216:2.5) node[left] {$\ul{2}$};
    \draw[gluon] (252:2.5) node[below] {$3$} -- (252:1.5);
  \end{tikzpicture},
\end{equation}
where we introduce the notation that quarks are labeled by underlined numbers and the corresponding antiquark with the same number overlined. Primitive amplitudes involving quarks are linearly dependent under a generalization of KK-relations, first observed in~\cite{Ita:2011ar} through non-trivial solutions of a linear system in terms of (Feynman) diagrams. Dyck words~\cite{Duchon:2000, Kasa:2010} allow for a compact description of a basis, i.e. a spanning set of linearly independent primitive amplitudes. A Dyck word is a list of opening and closing brackets, that are composed in a mathematically correct way. A pair of brackets corresponds to a quark-antiquark pair, i.e. we identify an opening (closing) bracket with a quark (antiquark). For $n$ external partons we denote the set of Dyck words including all permutations of $k$ quark-antiquark pairs and $n-2k$ insertions of gluons by $\text{Dyck}_{n,k}$.

Consider the Dyck word $\mlq()(()())\mrq$. For $n=12$ external states, including $k=4$ quark-antiquark pairs, we can for example make an assignment of partons of the form
\begin{equation}
  \overset{(}{\ul{1}\strut}2\overset{)}{\ol{1}\strut}3\overset{(}{\ul{4}\strut}\overset{(}{\ul{5}\strut}6\overset{)}{\ol{5}\strut}7\overset{(}{\ul{8}\strut}\overset{)}{\ol{8}\strut}\overset{)}{\ol{4}\strut} \in \text{Dyck}_{12,4}.
\end{equation}
We assume here and in the following that all quark-antiquark pairs have distinct flavor -- the one-flavor case can be recovered by summing over all combinations of quark-antiquark pairings (see e.g.~\cite{Reuschle:2013qna,Melia:2013epa}). The numbers 1, 4, 5 and 8 label quark-antiquark pairs, whereas 2, 3, 6 and 7 mark inserted gluons at various places. Note that the quark~$\underline{i}$ and the corresponding antiquark~$\overline{i}$ have different momenta even though they carry the same number. A basis of primitive amplitudes is then simply given by
\begin{equation}\label{eqn:basis_tree}
  \mathcal{B}^0_{n,k} = \left\{A^{(0)}(\ul{1},\sigma,\ol{1}) \mid \sigma \in \text{Dyck}_{n-2,k-1}\right\},
\end{equation}
where one quark-antiquark pair is fixed at the beginning and the end. The size of this basis has been shown to be $(n-2)!/k!$. This basis turns out to be useful to generalize the DDM color decomposition. We discuss the full tree level color decomposition of QCD in the next section.

\subsection{Color decomposition of QCD tree amplitudes}
A trace-based color decomposition for QCD for a $\text{SU}(N_c)$ gauge group leads to rather complex expressions~\cite{Ellis:2011cr, Ita:2011ar, Bern:1994fz, Badger:2012pg, Reuschle:2013qna, Schuster:2013aya}. The JO color decomposition~\cite{Johansson:2015oia} provides a structured form which holds for any number of external partons and for any gauge group.

Given Melia's basis of primitive amplitudes, the JO decomposition proposes a form of the  color factors $C^{(0)}$ such that the full amplitude (for a fixed number of external particles~$n$ with~$k$ quark-antiquark pairs) is recovered by
\begin{equation}
  \label{eq:tree_color_dec}
  \mathcal{A}^{(0)}_{n,k} = g^{n-2}\smashoperator{\sum_{\sigma\in\text{Dyck}_{n-2,k-1}}} C^{(0)}(\ul{1},\sigma,\ol{1}) A^{(0)}(\ul{1},\sigma,\ol{1}).
\end{equation}
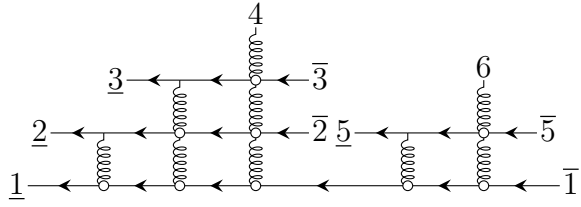
\begin{figure}
  \centering
  \input{tikzMarioWorlds}
  \caption{Example of a `Mario world' color diagram using JO's diagrammatic notation. The diagram describes the color factor $C^{(0)}(\ul{1},\ul{2},\ul{3},4,\ol{3},\ol{2},\ul{5},6,\ol{5},\ol{1})$. It corresponds to the primitive amplitude given in (\ref{eqn:ex_tree_prim}).}
  \label{fig:mario_worlds}
\end{figure}%
The result is most easily understood in a diagrammatic way. The color diagrams that contribute to a color factor have a particular structure, which has been named `Mario world' due to its similarity with the virtual world of a certain arcade game~\cite{mario}\footnote{We use the term Mario world diagram for a single color diagram as well as for diagrams representing a color factor (a sum of color diagrams) which we write using composite vertices as explained below.}. An example is given in fig.~\ref{fig:mario_worlds}. The color factor of a single graph is determined only considering the color part of the Feynman rules. We use the conventions
\begin{align}
  \tilde{f}^{abc} &=
  \begin{tikzpicture}
    [>=stealth, baseline=-0.5, scale=0.8]
    \draw[gluon] (-30:1) node[right] {$c$}-- (0,0);
    \draw[gluon] (90:1) node[above] {$b$} -- (0,0);
    \draw[gluon] (210:1) node[left] {$a$}-- (0,0);
  \end{tikzpicture},
  &
  T_{i\bar{\jmath}}^a &=
  \begin{tikzpicture}
    [>=stealth, baseline=-0.5, scale=0.8]
    \draw[fermion] (-30:1) node[right] {$\bar{\jmath}$}-- (0,0);
    \draw[gluon] (90:1) node[above] {$a$} -- (0,0);
    \draw[fermion] (0,0) -- (210:1) node[left] {$i$};
  \end{tikzpicture},
  &
  \widebar{T}_{\bar{\jmath}i}^a &=
  \begin{tikzpicture}
    [>=stealth, baseline=-0.5, scale=0.8]
    \draw[fermion] (0,0)-- (-30:1) node[right] {$i$};
    \draw[gluon] (90:1) node[above] {$a$} -- (0,0);
    \draw[fermion] (210:1) node[left] {$\bar{\jmath}$} -- (0,0);
  \end{tikzpicture}
  = - T_{i\bar{\jmath}}^a.
\end{align}
The notation of fundamental generators with flipped indices~$\widebar{T}^a_{\bar{\jmath}i} \equiv - T^a_{i\bar{\jmath}}$ is used to introduce an artificial  antisymmetry for the fundamental generators, similar to the adjoint generators~$(T^a_{\text{adj}})_{bc} \equiv \tilde{f}^{bac}$:
\begin{equation}
  \begin{aligned}
    \tilde{f}^{cab} &= - \tilde{f}^{bac},\\
    T^a_{\bar{\jmath}i} &= -T^a_{i\bar{\jmath}}.
  \end{aligned}
\end{equation}
Color factors of different diagrams are related by the Jacobi identity and the defining commutation relation, inherited from the color-algebra of the gauge group. The identities for the adjoint and the fundamental representation are given by
\begin{equation}
  \begin{aligned}
    \tilde{f}^{dac}\tilde{f}^{cbe} - \tilde{f}^{dbc}\tilde{f}^{cae} &= \tilde{f}^{abc}\tilde{f}^{dce},\\
    T^a_{i\bar{j}} T^b_{j\bar{k}} - T^b_{i\bar{j}} T^a_{j\bar{k}} &= \tilde{f}^{abc} T^c_{i\bar{k}}.
  \end{aligned}
\end{equation}
The relations in terms of diagrams are
\begin{align}
  \begin{tikzpicture}
    [>=stealth, baseline=18, scale=0.5]
    \draw[gluon] (0,0) -- (1,0.8);
    \draw[gluon] (2,0) -- (1,0.8);
    \draw[gluon] (1,0.8) -- (1,2.2);
    \draw[gluon] (0,3) -- (1,2.2);
    \draw[gluon] (2,3) -- (1,2.2);
  \end{tikzpicture}
  -
  \begin{tikzpicture}
    [>=stealth, baseline=18, scale=0.5]
    \draw[gluon] (-1,0) -- (1,2.4);
    \draw[gluon] (2,0) -- (1,0.6);
    \draw[gluon] (1,0.6) -- (1,2.4);
    \node[fill=white, circle, inner sep=3.2] at (0.25,1.5) {};
    \draw[gluon] (-1,3) -- (1,0.6);
    \draw[gluon] (2,3) -- (1,2.4);
  \end{tikzpicture}
  &=
  \begin{tikzpicture}
    [>=stealth, baseline=12, scale=0.5]
    \draw[gluon] (0,0) -- (0.8,1);
    \draw[gluon] (3,0) -- (2.2,1);
    \draw[gluon] (0.8,1) -- (2.2,1);
    \draw[gluon] (0,2) -- (0.8,1);
    \draw[gluon] (3,2) -- (2.2,1);
  \end{tikzpicture},\\
  \begin{tikzpicture}
    [>=stealth, baseline=18, scale=0.5]
    \draw[gluon] (0,0) -- (1,0.8);
    \draw[fermion] (1,0.8) -- (2,0);
    \draw[fermion] (1,2.2) -- (1,0.8);
    \draw[gluon] (0,3) -- (1,2.2);
    \draw[fermion] (2,3) -- (1,2.2);
  \end{tikzpicture}
  -
  \begin{tikzpicture}
    [>=stealth, baseline=18, scale=0.5]
    \draw[gluon] (-1,0) -- (1,2.4);
    \draw[fermion] (1,0.6) -- (2,0);
    \draw[fermion] (1,2.4) -- (1,0.6);
    \node[fill=white, circle, inner sep=3.2] at (0.25,1.5) {};
    \draw[gluon] (-1,3) -- (1,0.6);
    \draw[fermion] (2,3) -- (1,2.4);
  \end{tikzpicture}
  &=
  \begin{tikzpicture}
    [>=stealth, baseline=12, scale=0.5]
    \draw[gluon] (0,0) -- (0.8,1);
    \draw[fermion] (2.2,1) -- (3,0);
    \draw[gluon] (0.8,1) -- (2.2,1);
    \draw[gluon] (0,2) -- (0.8,1);
    \draw[fermion] (3,2) -- (2.2,1);
  \end{tikzpicture}.
\end{align}
These identities applied to subparts of a color diagram lead to linear relations between color factors. In our case the number of linearly independent color diagrams matches the size of the basis of primitive amplitudes for fixed number of external states~$n$ and number of quark-antiquark pairs~$k$. This means that this color decomposition is minimal in the sense that the sum~(\ref{eq:tree_color_dec}) is over a minimal set of (independent) color factors and primitive amplitudes and it is justified to call~(\ref{eqn:basis_tree}) a \emph{basis}.

The last ingredient needed is an operation on color diagrams introduced by JO. We present a mathematical as well as a diagrammatic notation to define the operator
\begin{align}
  \Xi_l^a &= \sum_{s=1}^l \underbrace{1\otimes \cdots \otimes 1 \otimes \overbrace{\cT^a \otimes 1 \otimes \cdots \otimes 1 \otimes 1}^s}_l\\
  &=
  \begin{tikzpicture}
    [>=stealth, baseline=30]
    \draw[fermion] (0.5,0) -- (0,0);
    \draw[fermion] (1,0) -- (0.5,0);
    \draw[fermion] (1,0.6) -- (0.5,0.6);
    \draw[fermion] (0.5,0.6) -- (0,0.6);
    \draw[fermion] (1,1.2) -- (0.5,1.2);
    \draw[fermion] (0.5,1.2) -- (0,1.2);
    \draw[fermion] (1,2) -- (0.5,2);
    \draw[fermion] (0.5,2) -- (0,2);
    \draw[decorate, decoration={brace, amplitude=6pt}] (1,2.1) -- node[right=4pt] {$l$} (1,-0.1);
    \draw[gluon] (0.5,0) -- (0.5,0.6);
    \draw[gluon] (0.5,0.6) -- (0.5,1.2);
    \draw[dotted] (0.5,1.3) -- (0.5,1.9);
    \draw[gluon] (0.5,2) -- (0.5,2.6) node[above] {$a$};
    \cC{(0.5,0)};
    \cC{(0.5,0.6)};
    \cC{(0.5,1.2)};
    \cC{(0.5,2)};
  \end{tikzpicture}
  =
  \begin{tikzpicture}
    [>=stealth, baseline=30]
    \draw[fermion] (0.5,0) -- (0,0);
    \draw[fermion] (1,0) -- (0.5,0);
    \draw[fermion] (1,0.6) -- (0.5,0.6);
    \draw[fermion] (0.5,0.6) -- (0,0.6);
    \draw[fermion] (1,1.2) -- (0.5,1.2);
    \draw[fermion] (0.5,1.2) -- (0,1.2);
    \draw[fermion] (1,2) -- (0.5,2);
    \draw[fermion] (0.5,2) -- (0,2);
    \node[fill=white, circle, inner sep=3] at (0.5,0.6) {};
    \node[fill=white, circle, inner sep=3] at (0.5,1.2) {};
    \node[fill=white, circle, inner sep=3] at (0.5,2) {};
    \draw[gluon] (0.5,0) -- (0.5,1.5);
    \draw[dotted] (0.5,1.5) -- (0.5,1.8);
    \draw[gluon] (0.5,1.8) -- (0.5,2.6);
  \end{tikzpicture}
  +
  \begin{tikzpicture}
    [>=stealth, baseline=30]
    \draw[fermion] (0.5,0) -- (0,0);
    \draw[fermion] (1,0) -- (0.5,0);
    \draw[fermion] (1,0.6) -- (0.5,0.6);
    \draw[fermion] (0.5,0.6) -- (0,0.6);
    \draw[fermion] (1,1.2) -- (0.5,1.2);
    \draw[fermion] (0.5,1.2) -- (0,1.2);
    \draw[fermion] (1,2) -- (0.5,2);
    \draw[fermion] (0.5,2) -- (0,2);
    \node[fill=white, circle, inner sep=3] at (0.5,1.2) {};
    \node[fill=white, circle, inner sep=3] at (0.5,2) {};
    \draw[gluon] (0.5,0.6) -- (0.5,1.5);
    \draw[dotted] (0.5,1.5) -- (0.5,1.8);
    \draw[gluon] (0.5,1.8) -- (0.5,2.6);
  \end{tikzpicture}
  +
  \begin{tikzpicture}
    [>=stealth, baseline=30]
    \draw[fermion] (0.5,0) -- (0,0);
    \draw[fermion] (1,0) -- (0.5,0);
    \draw[fermion] (1,0.6) -- (0.5,0.6);
    \draw[fermion] (0.5,0.6) -- (0,0.6);
    \draw[fermion] (1,1.2) -- (0.5,1.2);
    \draw[fermion] (0.5,1.2) -- (0,1.2);
    \draw[fermion] (1,2) -- (0.5,2);
    \draw[fermion] (0.5,2) -- (0,2);
    \node[fill=white, circle, inner sep=3] at (0.5,2) {};
    \draw[gluon] (0.5,1.2) -- (0.5,1.5);
    \draw[dotted] (0.5,1.5) -- (0.5,1.8);
    \draw[gluon] (0.5,1.8) -- (0.5,2.6);
  \end{tikzpicture}
  + \cdots +
  \begin{tikzpicture}
    [>=stealth, baseline=30]
    \draw[fermion] (0.5,0) -- (0,0);
    \draw[fermion] (1,0) -- (0.5,0);
    \draw[fermion] (1,0.6) -- (0.5,0.6);
    \draw[fermion] (0.5,0.6) -- (0,0.6);
    \draw[fermion] (1,1.2) -- (0.5,1.2);
    \draw[fermion] (0.5,1.2) -- (0,1.2);
    \draw[fermion] (1,2) -- (0.5,2);
    \draw[fermion] (0.5,2) -- (0,2);    
    \draw[dotted] (0.5,1.3) -- (0.5,1.9);
    \draw[gluon] (0.5,2) -- (0.5,2.6);
  \end{tikzpicture},
\end{align}
with $l$ the \emph{nestedness} of the gluon line (i.e. how many quark lines are below the given gluon line) with adjoint index~$a$. There is a difference to the notation used in~\cite{Johansson:2015oia}: We insert a general object $\cT^a$ which stands for either $T^a$ or $\widebar{T}^a$ depending on the orientation of the quark line this object connects to. This notation is independent of the chosen quark line directions and will be especially useful for the one-loop discussion. For our choice of quark line directions in the tree level case we will always have that $\cT^a = T^a$.\footnote{Note that~\cite{Johansson:2015oia} uses a different convention where the lowest quark line points in the opposite direction. They correspondingly insert $\widebar{T}^a$ at the lowest level $s=1$.}

The operator $\Xi$ allows for a compact definition of the color factors in the decomposition~(\ref{eq:tree_color_dec})
\begin{equation}
  C^{(0)}(\ul{1},\sigma,\ol{1}) = \{\ul{1}|\sigma|\ol{1}\} \bigg\rvert{\begin{subarray}{l}
      \ul{q} \rightarrow \{\ul{q}| \cT^b \otimes \Xi^b_{l-1},\\
      \ol{q} \rightarrow |\ol{q}\},\\
      g \rightarrow \Xi^{a_g}_{l}
  \end{subarray}},
\end{equation}
where we use the bracket notation to indicate fundamental color indices and $a_g$ is the adjoint index of the external gluon~$g$. Note that the nestedness of a quark line also includes the line itself. A precise definition of the object~$\cT^a$ introduced before, and the bracket notation, is
\begin{equation}
  \label{eq:bracket_notation}
  \begin{aligned}
    \{\ul{i}|\cT^{a_1} \cdots \cT^{a_m}|\ol{j}\} &= \{\ul{i}|T^{a_1} \cdots T^{a_m}|\ol{j}\} = (T^{a_1} \cdots T^{a_m})_{i\bar{j}},\\
    \{\ol{i}|\cT^{a_1} \cdots \cT^{a_m}|\ul{j}\} &= \{\ol{i}|\ol{T}^{a_1} \cdots \ol{T}^{a_m}|\ul{j}\} = (-1)^m (T^{a_m} \cdots T^{a_1})_{j\bar{i}},\\
  \end{aligned}
\end{equation}
which implements the correct contraction of fundamental indices depending on the orientation of the quark line. A bracket with nestedness~$l$ acts on the part of the tensor product with the same nestedness. The evaluation is most easily done by resolving brackets from inside to outside, using
\begin{equation}
  \label{eq:eval_tensor}
  \begin{aligned}
    &\{i|\cdots \{k|(\cT^{a_1} \cdots \cT^{a_r})\otimes (\cT^{b_1} \cdots \cT^{b_s}) \otimes \cdots |l\} \cdots |j\}\\
    =\,&\{i|\cdots \{k|\cT^{a_1} \cdots \cT^{a_r}|l\} (\cT^{b_1} \cdots \cT^{b_s}) \otimes \cdots |j\}\\
    =\,&\{k|\cT^{a_1} \cdots \cT^{a_r}|l\} \{i|\cdots (\cT^{b_1} \cdots \cT^{b_s}) \otimes \cdots |j\}
  \end{aligned}
\end{equation}
recursively.

Diagrammatically a color factor~$C^{(0)}(\ul{1},\sigma,\ol{1})$ is given by a Mario world diagram with base line $\ul{1} \leftarrow \ol{1}$, and external ordering and nesting according to the Dyck word~$\sigma$. The purely gluonic case (i.e. the DDM color decomposition) can be recovered by defining $\Xi^a_0 = T^a_{\text{adj}}$ that acts on the `zeroth level' of the Mario world diagram, i.e. simply contracting the appropriate $\Xi_0^a$ operators and fixing two gluons at the start and the end instead of a quark-antiquark pair.

A number of examples of this color decomposition can be found in the original paper~\cite{Johansson:2015oia}. This completes the QCD tree level color decomposition for any number of external gluons and quark-antiquark pairs. We discuss the one-loop extension in the rest of these notes.

\section{A basis for primitive one-loop amplitudes}\label{sec:basis}
We begin by defining primitive one-loop QCD amplitudes. These primitive amplitudes are planar and color-ordered. A basis of independent primitive amplitudes is constructed using Dyck words, and builds up a minimal color decomposition, given the corresponding color factors -- similar to the tree level results reviewed above.

We define primitive one-loop amplitudes $A_{n,k}$ following~\cite{Bern:1994fz}, where $n$ is the number of external partons and $k$ counts the quark-antiquark pairs, such that $n-2k$ is the number of external gluons. Primitive one-loop amplitudes are purely kinematic objects analogous to color-ordered tree level amplitudes. They are built from planar diagrams with a fixed (cyclic) ordering of external particles -- and for amplitudes including quarks (external or in the loop) a fixed routing of these elements, i.e. a primitive amplitude specifies on which side a quark line, a gluon or a loop lies with respect to all quark lines. Additionally we distinguish amplitudes that have a closed fermion loop from ones that have at least one gluon line carrying loop momentum. Primitive amplitudes can be directly computed using color-ordered Feynman rules (see e.g. \cite{Dixon:1996wi}), summing up all the planar graphs with the specified routing of quark lines. Modern methods like generalized unitary~\cite{Bern:1994zx, Bern:1994cg, Carrasco:2011hw, Bern:2011qt, Ita:2011hi, Britto:2010xq}, (loop-level) BCFW~\cite{Britto:2004ap, Britto:2005fq, NigelGlover:2008ur, Bierenbaum:2010cy, Elvang:2011ub, CaronHuot:2010zt, Boels:2010nw} or Q-cuts~\cite{Baadsgaard:2015twa, Huang:2015cwh}  simplify this computation for QCD as well as for supersymmetric extensions thereof.

Note that for massless theories diagrams with a bubble on an external leg or a tadpole of the form
\begin{equation*}
  \begin{tikzpicture}
    \draw (0:0.3) arc (0:360:0.3);
    \draw (0:0.3) -- (0:0.8);
  \end{tikzpicture}
\end{equation*}
come with an ill-defined vanishing propagator. From a purely diagrammatic point of view we observe that the inclusion of external bubbles and tadpoles with a closed quark loop pose no problem to the color-decomposition and can formally be included. A gauge group with vanishing traces of corresponding Lie algebra generators $\Tr(T^a)=0$\footnote{This constraint for example appears for theories whose gauge group contains factors of $U(1)$, for which it has been shown that they cannot be consistently coupled to gravity unless all $U(1)$ generators are traceless \cite{AlvarezGaume:1983ig}.}, e.g. a semisimple Lie algebra, induces vanishing color factors for diagrams containing a quark tadpole and so these contributions can be ignored.  In the following we will though mostly give attention to the more general case assuming the presence of diagrams with quark tadpoles, validating the discussion for any gauge group. In the case that the trace of every generator of the Lie algebra is zero the results obtained here can be slightly simplified as we will remark accordingly. For gluonic tadpoles the color decomposition of the form proposed no longer holds on a formal level. Imposing that the color factors of such diagrams vanish due to the antisymmetry of the structure constants fixes their behavior and we can safely ignore them. 

An example for a 10-point one-loop primitive amplitude can be represented as
\begin{equation}\label{eq:ex_primitive}
  A^{q^\tL[1^\tL,2^\tR,3^\tL,5^\tR]}(\ul{1},\ol{1},\ul{2},\ul{3},\ol{3},4,\ol{2},\ol{5},\ul{5},6) \, = \,
  \begin{tikzpicture}
    [>=stealth, baseline=-0.5, scale=0.7]
    \draw[dashed] (0:2) arc (0:360:2);
    \draw[fermion] (-36:2.5) node[right] {$\ol{1}$} .. controls (-36:1) and (0:1) .. (0:2.5) node[right] {$\ul{1}$};
    \draw[fermion] (-216:2.5) node[left] {$\ol{2}$} .. controls (-276:1) and (-2:1) .. (-72:2.5) node[below] {$\ul{2}$};
    \draw[fermion] (-144:2.5) node[left] {$\ol{3}$} .. controls (-144:1) and (-108:1) .. (-108:2.5) node[below] {$\ul{3}$};
    \draw[fermion] (-252:2.5) node[above] {$\ol{5}$} .. controls (-252:1) and (-288:1) .. (-288:2.5) node[above] {$\ul{5}$};
    \draw[gluon] (-180:2.5) node[left] {$4$} -- (-180:1.5);
    \draw[gluon] (-324:2.5) node[right] {$6$} -- (-324:1.5);
    \draw[fermion] (-0.05,-0.2) arc (0:-360:0.5);
  \end{tikzpicture}.
\end{equation}
The superscript $q^\tL[1^\tL,2^\tR,3^\tL,5^\tR]$ specifies the type of loop ($q$ for a closed quark loop or $g$ for a loop that contains at least one gluonic line) and its properties. If the loop is a quark loop the superscript $q^{\tL/\tR}$ specifies on which side the external particles lie with respect to the loop line direction -- equivalently one could specify the orientation of the loop. The rest of the arguments of the superscript specify on which side the quark loop lies with respect to the quark line with the given number, i.e. for this example the quark loop lies on the left side of first quark line, on the right side of the second loops line and so on. There is no need to specify the side on which each quark line or gluon lies with respect to all other quark lines since this is clear by the external ordering, if one considers only non-vanishing primitive amplitudes.

A primitive amplitude is computed by summing up all planar diagrams that have the given external ordering and quark line routing. In the language of the diagrammatic representation~(\ref{eq:ex_primitive}) this is achieved by drawing all possible planar graphs on the annulus that is bounded by the loop line on the inside and the dashed line carrying the external partons on the outside. Planarity allows for a uniform treatment of loop-momentum assignment~\cite{ArkaniHamed:2010gh}. Using dual-space coordinates ${p_i \equiv x_i - x_{i-1}}$~\cite{Drummond:2006rz} one defines the loop momentum via $\ell \equiv x_0 - x_n$ which corresponds to the convention that the loop momentum flows between the last and the first external leg on the annulus
\begin{equation}
  \begin{tikzpicture}
    [>=stealth, baseline=-0.5, scale=0.7]
    \draw[dashed] (0:1) arc (0:360:1);
    \draw[dashed] (0:2) arc (0:360:2);
    \draw[dotted] (0:0.5) -- (0:2.5) node[right] {$x_n$};
    \node at (0,0) {$x_0$};
    \node at (90:2.5) {$x_{n-1}$};
    \node at (-90:2.5) {$x_1$};
    \draw[fermion] (45:1.8) -- (45:2.5) node[above right] {$p_n$};
    \draw[fermion] (-45:1.8) -- (-45:2.5) node[below right] {$p_1$};
    \draw[fermion] (135:1.8) -- (135:2.5) node[above left] {$p_{n-1}$};
    \draw[fermion] (-135:1.8) -- (-135:2.5) node[below left] {$p_{2}$};
    \draw[dotted] (160:2.9) arc (160:200:2.9);
    \draw[fermion] (30:1.3) arc (30:-30:1.3) node[midway,right] {$\ell$};
  \end{tikzpicture}.
\end{equation}  

From all of these primitive amplitudes it is possible to pick a subset, that we will call a basis, in the sense that they form a maximal set of linearly independent elements, and we can express the full color-dressed amplitudes in terms of these objects and color factors only. Linear relations between primitive amplitudes can be found by expressing primitive amplitudes in terms of planar diagrams -- assuming that distinct diagrams are independent -- and reducing the emerging system of equations. This general algorithm to find a set of linearly independent primitive amplitudes is described in detail in~\cite{Ita:2011ar}. A basis obtained in this way is in general not unique.

It turns out that the number of linearly independent primitive amplitudes is the same as the number of independent (with respect to Jacobi identities and commutation relations) color factors of diagrams of the same process as expected for a basis of kinematic objects. In this sense the color decomposition presented below is over a minimal set of primitive amplitudes and color factors as in the tree level case. We check the matching of these numbers explicitly up to at least seven points for all possible combinations of external particles and every loop type. Note that for a gauge group with traceless algebra generators $\Tr(T^a)=0$ the number of independent color factors is reduced since diagrams containing a quark tadpole have a vanishing color structure. The number of independent primitive amplitudes shrinks by the same amount when diagrams with a tadpole that is formed by a quark line -- corresponding to exactly these vanishing color factors -- are ignored.

We discuss our choice of basis for the case of a closed quark loop separately from the case of a mixed or purely gluonic loop. The special case of a purely gluonic amplitude we will treat separately. Note that the above example~(\ref{eq:ex_primitive}) of a primitive amplitude is not a basis element for our choice of basis.

\subsection{Closed quark loop}
As for the case of tree level amplitudes, a basis for primitive one-loop amplitudes is most conveniently presented using Dyck words~\cite{Melia:2013bta} and the Mario world structure. In contrast to the tree level case one does not fix a quark line as the base of the Mario world diagram, instead the quark loop plays the role of this line. Since we do not fix any external particles we need the notion of cyclicity for Dyck words. We define the set of cyclically distinct Dyck words including gluon insertions and permutations of quark-antiquark pairs by
\begin{equation}
  \text{Dyck}_{n,k}^\circlearrowleft = \text{Dyck}_{n,k}/\{\text{cyclic transformations}\}.
\end{equation}
A basis element is represented by such a cyclic Dyck word. For example, a possible assignment of quark line numbers and insertions of gluons for the Dyck word `$(())()$' is $\ul{1}\ul{2}3\ol{2}\ol{1}\ul{4}5\ol{4}\in\text{Dyck}_{n,k}^\circlearrowleft$ which represents a primitive amplitude that diagrammatically has the form
\begin{equation}
  \label{eq:ex_primitive_quark}
  A^{q^\tR[1^\tL2^\tL4^\tL]}(\ul{1}\ul{2}3\ol{2}\ol{1}\ul{4}5\ol{4}) \, = \,
  \begin{tikzpicture}
    [>=stealth, baseline=-0.5, scale=0.7]
    \draw[dashed] (0:2) arc (0:360:2);
    \draw[fermion] (-180:2.5) node[left] {$\ol{1}$} .. controls (-130:1) and (-50:1) .. (0:2.5) node[right] {$\ul{1}$};
    \draw[fermion] (-135:2.5) node[left] {$\ol{2}$} .. controls (-135:1) and (-45:1) .. (-45:2.5) node[right] {$\ul{2}$};
    \draw[fermion] (-315:2.5) node[right] {$\ol{4}$} .. controls (-315:1) and (-225:1) .. (-225:2.5) node[left] {$\ul{4}$};
    \draw[gluon] (-90:2.5) node[below] {$3$} -- (-90:1.5);
    \draw[gluon] (-270:2.5) node[above] {$5$} -- (-270:1.5);
    \draw[fermion] (0.5,0.2) arc (0:360:0.5);
  \end{tikzpicture},
\end{equation}
where 1, 2 and 4 label quark lines, and 3 and 5 gluon insertions. We choose the quark loop to lie on the left side of every quark line, and every quark line to the right of the loop line. As will be clear when discussing the color decomposition, this choice of quark routings allows us to recover the tree level conventions when cutting through the quark loop. This information fully specifies a primitive amplitude.

A complete basis for fixed~$n$ and~$k$ is given by all cyclic Dyck words:
\begin{equation}
  \label{eq:basis_quark}
  \mathcal{B}^q_{n,k} = \left\{ A^{q^L[i^L \mid i \in \{\text{quark lines}\}]}(\sigma) \mid \sigma \in \text{Dyck}_{n,k}^\circlearrowleft \right\}.
\end{equation}
The size of this basis can be computed using a recursive description of Dyck words. The number of Dyck words with $k$ brackets without any labeling or gluon insertion is given by the Catalan number~$C_k = (2k)!/((k+1)!k!)$. In order to keep track of cyclicity we demand that the first quark is always assigned to the first bracket or to one of the brackets inside the first closed substructure (i.e. between the first opening and its corresponding closing bracket). An explicit expression for the number of labeled cyclic Dyck words (without gluon insertions) is the number of all possible Dyck words inside the first closed substructure~$C_s$ of length~$s$, times the number of (non-cyclic) Dyck words this structure can be followed by~$C_{k-s-1}$. Including the permutation of the all except the first quark-antiquark pair~$(k-1)!$, this statement is expressed as
\begin{equation}
  \lvert \text{Dyck}_{2k,k}^\circlearrowleft\rvert = \sum_{s=0}^{k-1} C_s C_{k-s-1} (s+1) (k-1)! = \frac{(2k-1)!}{k!},
\end{equation}
where the factor $(s+1)$ counts all possible assignments of the first quark-antiquark pair inside the first closed substructure. The number of available slots for the first gluon insertion is given by $2k$ and is enlarged by one for every further gluon. The size of the basis is then computed to be
\begin{equation}
  \lvert \text{Dyck}_{n,k}^\circlearrowleft \rvert = \frac{(2k-1)!}{k!}(2k)(2k+1)\cdots(n-1) = \frac{(n-1)!}{k!}.
\end{equation}
The size of the basis is the same as the number of independent color factors, as we check explicitly for the cases listed in table~\ref{tbl:checks_quark}. 
\begin{table}
  \centering
  \begin{tabular}{l | c c c c c c c c c c}
    $k \setminus n$ & 2 & 3 & 4 & 5 & 6 & 7 & 8 & 9 & 10\\
    \midrule
    0 & 1 & 2 & 6 & 24 & 120 & 720\\
    1 & 1 & 2 & 6 & 24 & 120 & 720\\
    2 & - & - & 3 & 12 & 60 & 360 & 2520\\
    3 & - & - & - & - & 20 & 120 & 840\\
    4 & - & - & - & - & - & - & 210 & 1680\\
    5 & - & - & - & - & - & - & - & - & 3024
  \end{tabular}
  \caption{This table lists the size of the basis of primitive amplitudes, or equivalently the number of independent color factors for amplitudes with a closed quark loop. For the cases listed explicit checks have been done for the size of the basis, the number of independent color factors and the color decomposition. The variable~$n$ counts the total number of external particles and~$k$ denotes the number of quark-antiquark pairs. The numbers are explicitly given by~$(n-1)!/k!$.}
  \label{tbl:checks_quark}
\end{table}

\begin{sloppypar}
  For the case of a gauge group with traceless Lie algebra generators the size of the basis is reduced to ${(n-2)!(n-k-1)/k!}$, since we ignore contributions from diagrams with quark tadpoles. A basis is obtained by removing primitive amplitudes where all externals lie `inside' a single quark-antiquark pair, i.e. primitive amplitudes corresponding to Dyck words of the form `$(\cdots)$', where the last bracket closes the first one. This prescription can directly be applied in the discussion of the color decomposition below to obtain a formula for a decomposition that is minimal for this case.
\end{sloppypar}

\subsection{Gluonic or mixed loop}
We first treat the case where at least one external quark-antiquark pair is present. As in the closed quark loop case we start with the notion of cyclic Dyck words, with the difference that one has to additionally consider a modified scheme for the assignment of quark/antiquark to opening/closing brackets. We use square brackets to indicate quark-antiquark pairs with inverted assignment of quark and antiquark. We use the conventions that we always flip the assignment of all quark lines of a whole substructure of a Dyck word starting from the outermost level. Consider for example `(()())[[[]]]()' where we flipped the quark-antiquark assignment for every bracket in the second substructure. 

A specific basis element is denoted by such a modified cyclic Dyck word. For example, given the modified Dyck word `$()[[]]$' with $n=8$ and $k=3$ we can assign/insert particle labels in the following way:
\begin{equation}
  \label{eq:ex_primitive_gluon}
  A^{g[1^\tL2^\tR3^\tR]}(\overset{(}{\ul{1}\strut}\overset{)}{\ol{1}\strut}\overset{[}{\ol{2}\strut}\overset{[}{\ol{3}\strut}4\overset{]}{\ul{3}\strut}\overset{]}{\ul{2}\strut}5) =
  \begin{tikzpicture}
    [>=stealth, baseline=-0.5, scale=0.72]
    \draw[dashed] (0:2) arc (0:360:2);
    \draw[fermion] (-45:2.5) node[right] {$\ol{1}$} .. controls (-45:1) and (-0:1) .. (0:2.5) node[right] {$\ul{1}$};
    \draw[fermion] (-90:2.5) node[below=-0.1] {$\ol{2}$} .. controls (-130:1) and (-230:1) .. (-270:2.5) node[above=-0.1] {$\ul{2}$};
    \draw[fermion] (-135:2.5) node[left] {$\ol{3}$} .. controls (-135:1) and (-225:1) .. (-225:2.5) node[left] {$\ul{3}$};
    \draw[gluon] (-180:2.5) node[left] {$4$} -- (-180:1.5);
    \draw[gluon] (-315:2.5) node[right] {$5$} -- (-315:1.5);
    \draw[gluon] (0.7,0.15) arc (0:-380:0.4);
  \end{tikzpicture},
\end{equation}
where we again fix the loop to lie on the left hand side of normal-directed quark lines and on the right hand side of inverted quark lines. In this example the quark lines with label~2 and~3 are inverted in the sense that the antiquark appears before the quark in the Dyck word.

A possible choice of basis for primitive amplitudes is then given by all modified cyclic Dyck words, where we restrict the modified Dyck words to never have a square bracket in the start (i.e. we do not invert the directions of quark lines in the first substructure). We call the corresponding set~$\text{mDyck}_{n,k}^\circlearrowleft$ such that we can write the basis as
\begin{equation}
  \mathcal{B}^g_{n,k>0} = \left\{ A^{g[r(\sigma)]}(\sigma) \left\lvert
  \sigma \in \text{mDyck}_{n,k}^\circlearrowleft
  \right.\right\},
\end{equation}
where $r$ represents the quark line routing that the modified Dyck word~$\sigma$ induces. The basis contains $2^{2k-1}(n-1)!k!/(2k)!$ elements, where again $n$ is the total number of external partons and $k$ is the number of quark-antiquark pairs. This can be seen by considering the number of (unlabeled) Dyck words with possible reflection of \emph{all} substructures~$z(k)$, which is determined by the recursion relation
\begin{equation}
  \begin{aligned}
    z(0) &= 1,\\
    z(k) &= 2\sum_{s=0}^{k-1} C_s z(k-s-1),
  \end{aligned}
\end{equation}
where $C_s$ is the Catalan number that counts the number of Dyck words. The recursion works as follows: We count all possible Dyck words inside the first closed substructure times all possible Dyck words with reflections that can follow. The recursion relation is solved by $z(k) = \binom{2k}{k}$ as one can check explicitly.

To find the size of the basis we do a similar counting as in the quark case, replacing the factor~$C_{k-s-1}$ by $z(k-s-1)$:
\begin{equation}
  \begin{aligned}
    \lvert \text{mDyck}_{n,k}^\circlearrowleft \rvert &= \left(\sum_{s=0}^{k-1} C_s z(k-s-1)(s+1)\right)(k-1)!\underbrace{(2k)(2k+1)\cdots(n-1)}_{\text{gluon insertions}}\\
    &= \frac{2^{2k-1}(n-1)!k!}{(2k)!}.
  \end{aligned}
\end{equation}

The special case where no external quark-antiquark pair is present has already been described in~\cite{DelDuca:1999rs}. In that case the basis consists of all permutations of the external gluons up to cyclicity, and additionally reflection
\begin{equation}
  \mathcal{B}^g_{n,0} = \left\{ A^g(1, \sigma) \mid \sigma \in S_{n-1}/\mathbb{Z}_2 \right\}.
\end{equation}
Again, the size of the basis is the same as the number of independent color factors for all combinations of external partons. We explicitly check this for the cases listed in table~\ref{tbl:checks_mixed}. We also note that the substructures, i.e. a closed subword at top level of the Dyck word, are tree-level basis elements, up to a possible global inversion of the quark lines inside the whole substructure. 

\begin{table}
  \centering
  \begin{tabular}{l | c c c c c c c c c }
    $k \setminus n$ & 2 & 3 & 4 & 5 & 6 & 7 & 8 & 9 & 10\\
    \midrule
    0 & 1 & 1 & 3 & 12 & 60 & 360 \\
    1 & 1 & 2 & 6 & 24 & 120 & 720\\
    2 & - & - & 4 & 16 & 80 & 480 & 3360\\
    3 & - & - & - & - & 32 & 192 & 1344\\
    4 & - & - & - & - & - & - & 384 & 3072\\
    5 & - & - & - & - & - & - & - & - & 6144
  \end{tabular}
  \caption{This table lists the size of the basis of primitive amplitudes, or equivalently the number of independent color factors for amplitudes with a gluonic or mixed loop. For the cases listed explicit checks have been done for the size of the basis, the number of independent color factors and the color decomposition. The variable~$n$ counts the total number of external particles and~$k$ denotes the number of quark-antiquark pairs. The numbers are explicitly given by~$2^{2k-1}(n-1)!k!/(2k)!$.}
  \label{tbl:checks_mixed}
\end{table}

The primitive amplitudes appearing in the two bases are sufficient to write down a color decomposition of any one-loop amplitude. In the next section we describe a decomposition that includes exactly these bases and a set of linearly independent color factors. From the independence of the color factors in the decomposition it follows that in anomaly-free theories the primitive amplitudes in the basis are gauge-invariant quantities. Since one can make different choices of fermion routings we conclude that any planar primitive amplitude is gauge-invariant.

\section{One-loop color decomposition}\label{sec:decomp}

This section presents the main results of this work, a conjecture for the color decomposition of one-loop QCD amplitudes. The decomposition uses the basis above and is minimal in the sense that the size of the basis is exactly the same as the number of independent color factors. We treat the case of a closed quark loop again separately from the rest. The full amplitudes for a fixed number of external particles~$n$, whereof~$k$ are quark-antiquark pairs, is then obtained by summing up contributions from both cases
\begin{equation}
  \mathcal{A}^{(1)}_{n,k} = \mathcal{A}^{q}_{n,k} + \mathcal{A}^g_{n,k},
\end{equation}
where the superscript $q$ denotes contributions with a closed quark loops and $g$ contributions with a mixed or purely gluonic loop.

\subsection{Color decomposition for a closed quark loop}

The only missing part for the color decomposition is the definition of the color factors associated to an element in the basis of primitive amplitudes that will allow us to write
\begin{equation}
  \label{eq:color_dec_quark}
  \mathcal{A}_{n,k}^q = g^n\smashoperator{\sum_{\sigma\in \text{Dyck}_{n,k}^\circlearrowleft}} C^q(\sigma) A^q(\sigma).
\end{equation}
We drop the superscript for the quark line routings since these are fixed by the conventions described in the above section. For this case the color factors can be obtained by a simple modification of the tree level prescription: In contrast to the tree level case the quark loop plays the role of the fixed external quark-antiquark pair at the base of the Mario world diagram. For example, a graphical representation of the color factor associated to the primitive amplitude in eq.~(\ref{eq:ex_primitive_quark}) is given by
\begin{equation}
  \label{eq:ex_color_dec_quark}
  C^q(\ul{1}\ul{2}3\ol{2}\ol{1}\ul{4}5\ol{4}) = \input{tikzExClosedQuark}.
\end{equation}
Formally the color factors are conveniently written using the bracket notation from the tree level case
\begin{equation}
  C^q(\sigma) = \{\ul{i}|\sigma|\ol{i}\} \bigg\rvert{\begin{subarray}{l}
      \ul{q} \rightarrow \{\ul{q}| \cT^b \otimes \Xi^b_{l-1},\\
      \ol{q} \rightarrow |\ol{q}\},\\
      g \rightarrow \Xi^{a_g}_{l}
  \end{subarray}},
\end{equation}
where we implicitly sum over the fundamental color index~$i$. Note that $l$ counts the nestedness of the parton including the loop line.

As an example, we work out the color factor~(\ref{eq:ex_color_dec_quark}), dropping the gluon line with label 3 for simplicity. First we plug in the replacement rules and expand the $\Xi$ operators
\begin{align}
  \MoveEqLeft[1] C^q(\ul{1}\ul{2}\ol{2}\ol{1}\ul{4}5\ol{4})\notag\\
  &= \{\ul{i}|\{\ul{1}|\cT^b\otimes\Xi^b_1\{\ul{2}|(\cT^c\otimes\Xi^c_2)|\ol{2}\}|\ol{1}\}
  \{\ul{4}|(\cT^d\otimes\Xi^d_1)\Xi^{a_5}_2|\ol{4}\}|\ol{i}\}\notag\\
  &\begin{multlined}[13cm]
     =\{\ul{i}|\{\ul{1}|\cT^b\otimes\cT^b\{\ul{2}|\cT^c\otimes(1\otimes\cT^c + \cT^c\otimes 1)|\ol{2}\}|\ol{1}\}\times\\
     \{\ul{4}|(\cT^d\otimes\cT^d)(1\otimes \cT^{a_5} + \cT^{a_5}\otimes 1)|\ol{4}\}|\ol{i}\}
   \end{multlined}\\
  &\begin{multlined}[b][13cm]
     =\{\ul{i}|\{\ul{1}|\cT^b\otimes\cT^b\{\ul{2}|\cT^c\otimes 1\otimes\cT^c + \cT^c\otimes\cT^c\otimes1|\ol{2}\}|\ol{1}\}\times\\
     \{\ul{4}|\cT^d\otimes\cT^d\cT^{a_5} + \cT^d\cT^{a_5}\otimes\cT^d|\ol{4}\}|\ol{i}\}
   \end{multlined}.\notag
\end{align}
In the next step we use the evaluation rule for the tensor product and the fundamental brackets~(\ref{eq:eval_tensor}) to arrive at
\begin{align}
  \MoveEqLeft[0.3] C^q(\ul{1}\ul{2}\ol{2}\ol{1}\ul{4}5\ol{4})\notag\\
  &\begin{multlined}[13cm]
     =\{\ul{i}|\{\ul{2}|\cT^c|\ol{2}\}\{\ul{1}|\cT^b\otimes\cT^b\cT^c +\cT^b\cT^c\otimes \cT^b|\ol{1}\}\times\\
     \left(\{\ul{4}|\cT^d|\ol{4}\}\cT^d\cT^{a_5} + \{\ul{4}|\cT^d\cT^{a_5}|\ol{4}\}\cT^d\right)|\ol{i}\}
   \end{multlined}\notag\\
  &\begin{multlined}[13cm]
     =\{\ul{2}|\cT^c|\ol{2}\}\{\ul{i}|\Big(\{\ul{1}|\cT^b|\ol{1}\}\cT^b\cT^c + \{\ul{1}|\cT^b\cT^c|\ol{1}\}\cT^b\Big)\times\\
     \Big(\{\ul{4}|\cT^d|\ol{4}\}\cT^d\cT^{a_5} + \{\ul{4}|\cT^d\cT^{a_5}|\ol{4}\}\cT^d\Big)|\ol{i}\}
   \end{multlined}\\
  &\begin{multlined}
     =\{\ul{2}|\cT^c|\ol{2}\}\times\\
     \Big(\{\ul{1}|\cT^b|\ol{1}\}\{\ul{4}|\cT^d|\ol{4}\}\{\ul{i}|\cT^b\cT^c\cT^d\cT^{a_5}|\ol{i}\}
     +\{\ul{1}|\cT^b|\ol{1}\}\{\ul{4}|\cT^d\cT^{a_5}|\ol{4}\}\{\ul{i}|\cT^b\cT^c\cT^d|\ol{i}\}\\
     +\{\ul{1}|\cT^b\cT^c|\ol{1}\}\{\ul{4}|\cT^d|\ol{4}\}\{\ul{i}|\cT^b\cT^d\cT^{a_5}|\ol{i}\}
     +\{\ul{1}|\cT^b\cT^c|\ol{1}\}\{\ul{4}|\cT^d\cT^{a_5}|\ol{4}\}\{\ul{i}|\cT^b\cT^d|\ol{i}\}\Big).
   \end{multlined}\notag
\end{align}
Using~(\ref{eq:bracket_notation}) we can write the result in a more convenient notation
\begin{align}
  \MoveEqLeft[1] C^q(\ul{1}\ul{2}\ol{2}\ol{1}\ul{4}5\ol{4})\notag\\
  &\begin{multlined}[t][13cm]
     =(T^c)_{i_2\ol{\imath}_2}
     \Big(T^b_{i_1\ol{\imath}_1} T^d_{i_4\ol{\imath}_4} \Tr(T^bT^cT^dT^{a_5})
     +T^b_{i_1\ol{\imath}_i} (T^dT^{a_5})_{i_4\ol{\imath}_4} \Tr(T^bT^cT^d)\\
     +(T^b\cT^c)_{i_1\ol{\imath}_1} (T^d)_{i_4\ol{\imath}_4} \Tr(T^bT^dT^{a_5})
     +(T^bT^c)_{i_1\ol{\imath}_1} (T^dT^{a_5})_{i_4\ol{\imath}_4} \Tr(T^bT^d)\Big),
   \end{multlined}
\end{align}
where the traces appear due to the summation over the fundamental index~$i$. This expression is valid for any choice of the gauge group. A similar treatment can be done for every color factor appearing in the decomposition. A Mathematica implementation of this algorithm is provided in an ancillary file.

Note that cutting through the quark loop of the Mario world representation of a color factor directly recovers a tree level color factor with $n+2$ external particles including $k+1$ quark-antiquark pairs. Conversely, closing the base quark line of a tree level color factor produces a loop-level color object.

\subsection{Color decomposition for a mixed or gluonic loop}

We first discuss the case where at least one quark-antiquark pair is present. As before we write the color decomposition as
\begin{equation}
  \label{eq:decomp_gluon}
  \mathcal{A}^g_{n,k>0} = g^n\smashoperator{\sum_{\sigma\in\text{mDyck}_{n,k}^\circlearrowleft}} C^g(\sigma)A^g(\sigma).
\end{equation}
The color factors $C^g(\sigma)$ can be given in a recursive way, reusing results from the tree level case. The graphical notation allows for a short description: The outermost quark lines in the Dyck word are integrated into the mixed loop. Inside of each of these quark lines the color factor takes the form of a tree-level color factor. Gluons that are in the outermost level are directly connected to the loop. Consider the primitive amplitude in eq. (\ref{eq:ex_primitive_gluon}). With the given prescription one can immediately write down the corresponding color factor in a diagrammatic form
\begin{equation}\label{eq:ex_color_factor_gluon}
  C^g(\ul{1}\ol{1}\ol{2}\ol{3}4\ul{3}\ul{2}5) = \input{tikzExMixed}.
\end{equation}
To find an expression for the color factors in terms of the tensor product structure as before we make use of the `zeroth level' of the tensor product that we introduced for the purely gluonic tree level case: $\Xi_0^a = T^a_{\text{adj}}$. We introduce replacement rules that will allow us to formally define the color factors
\begin{equation}
  \label{eq:repl_rules}
  \mathcal{R} = \left\{
  \begin{aligned}
    q^\text{opening} &\rightarrow \{q| \cT^b \otimes \Xi_{l-1}^b, &l > 1 \\
    q^\text{opening} &\rightarrow |b\}\{q| \cT^b, &l = 1 \\
    q^\text{closing} &\rightarrow |q\}, &l > 1\\
    q^\text{closing} &\rightarrow \cT^b |q\}\{b| &l = 1\\
    g &\rightarrow \Xi^{b_g}_{l}
  \end{aligned}
  \right.,
\end{equation}
where $q^\text{opening}$ stands for either a quark or an antiquark that corresponds to a opening bracket and similar for the `closing' superscript. The outermost quark lines receive a contribution to the level zero tensor product using the bracket notation with an adjoint index~$b$. Note that the replacement rules for quarks at nestedness level $l=1$ should in principle receive an additional minus sign since the gluon line connects from below. This minus sign cancels out between the opening and closing bracket for every quark-antiquark pair at this level.

The formula for the color factors then simply reads
\begin{equation}
  \label{eq:color_gluon}
  C^g(\sigma) = \{a| \sigma |a\}\rvert_\mathcal{R},
\end{equation}
where a sum over the adjoint index~$a$ of the gluon loop line is implicit. Brackets with adjoint indices are evaluated using the rules
\begin{equation}
  \begin{aligned}
    \{a||b\} &= \delta^{ab},\\
    \{a|T^b_{\text{adj}}|c\} &= \tilde{f}^{abc},
  \end{aligned}
\end{equation}
and extensions thereof for more insertions for $T_\text{adj}$. To exemplify this rule we explicitly work out the above color factor~(\ref{eq:ex_color_factor_gluon}). First we use the replacement rules~(\ref{eq:repl_rules}) and expand the tensor product inside the $\Xi$ operators
\begin{align}
  \MoveEqLeft[1] C^g(\ul{1}\ol{1}\ol{2}\ol{3}4\ul{3}\ul{2}5)\notag\\
    &=\{a||b\}\{\ul{1}|\cT^b\cT^c|\ol{1}\}\{c||d\}\{\ol{2}|\cT^d
    \{\ol{3}|(\cT^e\otimes\Xi^e_1)\Xi^{a_4}_2|\ul{3}\}
    \cT^f|\ul{2}\}\{f|\Xi^{a_5}_0|a\}\notag\\
    &= \delta^{ab}\{\ul{1}|\cT^b\cT^c|\ol{1}\}\delta^{cd}
    \{\ol{2}|\cT^d\{\ol{3}|(\cT^e\otimes\cT^e)(1\otimes \cT^{a_4} + \cT^{a_4}\otimes 1)|\ul{3}\}
    \cT^f|\ul{2}\}\{f|T^{a_5}_{\text{adj}}|a\}\notag\\
    &= \delta^{ab}\delta^{cd}\{\ul{1}|\cT^b\cT^c|\ol{1}\}
    \{\ol{2}|\cT^d
    \{\ol{3}|\cT^e\otimes\cT^e\cT^{a_4} + \cT^e\cT^{a_4}\otimes\cT^e|\ul{3}\}\cT^f|2\}\tilde{f}^{fa_5a}.
\end{align}
We can now use the evaluation rule for the tensor product and the fundamental brackets~(\ref{eq:eval_tensor})
\begin{align}
  \MoveEqLeft[1] C^g(\ul{1}\ol{1}\ol{2}\ol{3}4\ul{3}\ul{2}5)\notag\\
    &=\{\ul{1}|\cT^a\cT^c|\ol{1}\}
    \{\ol{2}|\{\ol{3}|\cT^e|\ul{3}\}\cT^c\cT^e\cT^{a_4}\cT^f+\{\ol{3}|\cT^e\cT^{a_4}|\ul{3}\}\cT^c\cT^e\cT^f|\ul{2}\}\tilde{f}^{fa_5a}\\
    &=\{\ul{1}|\cT^a\cT^c|\ol{1}\}
    \big(\{\ol{2}|\cT^c\cT^e\cT^{a_4}\cT^f|\ul{2}\}\{\ol{3}|\cT^e|\ul{3}\}
    + \{\ol{2}|\cT^c\cT^e\cT^f|\ul{2}\}\{\ol{3}|\cT^e\cT^{a_4}|\ul{3}\}\big)\tilde{f}^{fa_5a}.\notag
\end{align}
We use~(\ref{eq:bracket_notation}), replacing $\cT^a \rightarrow \widebar{T}^a$ for the quark
lines~2 and~3 with inverted direction
\begin{align}
  \MoveEqLeft[1] C^g(\ul{1}\ol{1}\ol{2}\ol{3}4\ul{3}\ul{2}5)\notag\\
    &=\{\ul{1}|T^aT^c|\ol{1}\}
    \big(\{\ol{2}|\widebar{T}^c\widebar{T}^e\widebar{T}^{a_4}\widebar{T}^f|\ul{2}\}\{\ol{3}|\widebar{T}^e|\ul{3}\}
    + \{\ol{2}|\widebar{T}^c\widebar{T}^e\widebar{T}^f|\ul{2}\}\{\ol{3}|\widebar{T}^e\widebar{T}^{a_4}|\ul{3}\}\big)\tilde{f}^{fa_5a}\notag\\
    &= -(T^aT^c)_{i_1\ol{\imath}_1}\big((T^fT^{a_4}T^eT^c)_{i_2\ol{\imath}_2}T^e_{i_3\ol{\imath}_3}
    + (T^fT^eT^c)_{i_2\ol{\imath}_2}(T^{a_4}T^d)_{i_3\ol{\imath}_3}\big)\tilde{f}^{fa_5a}.
\end{align}
This result is valid for any choice of gauge group. A Mathematica implementation of this algorithm is provided in an ancillary file.

The color decomposition for the special case of a purely gluonic amplitude~\cite{DelDuca:1999iql} is obtained by changing the sum in~(\ref{eq:decomp_gluon}) to
\begin{equation}
  \mathcal{A}^g_{n>2,0} = g^n\smashoperator{\sum_{\sigma\in S_{n-1}(\{2,\dots,n\})/\mathbb{Z}_2}}C^g(1,\sigma)A^g(1,\sigma).
\end{equation}
For the very special case of the two-point purely gluonic amplitude the color decomposition comes with an additional factor~$1/2$
\begin{equation}
  \mathcal{A}^g_{2,0} = \frac{g^2}{2} C^g(1,2)A^g(1,2).
\end{equation}
The factor $1/2$ is a relict from the symmetry factor of the only contributing diagram. The definition of color factors in the purely gluonic case is the same as for the generic case.

\subsection{Example: $n=5$, $k=2$}\label{sec:ex}
We discuss a complete example for the color decomposition of a 5-point, one-loop amplitude with two quark-antiquark pairs, labeled by 1 and 2, and an external gluon with number 3. For the two types of loop we will first work out the basis and then write down the corresponding color factors -- using the graphical notation -- that make up the full color decomposition

Let us give the basis of primitive amplitudes in the case of in internal closed quark loop according to the prescription~(\ref{eq:basis_quark}) in terms of labeled cyclic Dyck words with the insertion of a single gluon:
\begin{equation}
  \label{eq:ex_quark_basis}
  \begin{aligned}
    \mathcal{B}^q_{5,2} = \big\{&A^q(\ul{1}3\ol{1}\ul{2}\ol{2}), A^q(\ul{1}\ol{1}3\ul{2}\ol{2}), A^q(\ul{1}\ol{1}\ul{2}3\ol{2}), A^q(\ul{1}\ol{1}\ul{2}\ol{2}3),\\
    & A^q(\ul{1}3\ul{2}\ol{2}\ol{1}), A^q(\ul{1}\ul{2}3\ol{2}\ol{1}), A^q(\ul{1}\ul{2}\ol{2}3\ol{1}), A^q(\ul{1}\ul{2}\ol{2}\ol{1}3),\\
    & A^q(\ul{2}3\ul{1}\ol{1}\ol{2}), A^q(\ul{2}\ul{1}3\ol{1}\ol{2}), A^q(\ul{2}\ul{1}\ol{1}3\ol{2}), A^q(\ul{2}\ul{1}\ol{1}\ol{2}3)\big\}.
  \end{aligned}
\end{equation}
where we organized the quark lines such that every line has a fixed quark structure and contains all four possible insertions of the gluon. We also dropped the superscript that encodes the quark routing since it is the same for all of these amplitudes. The corresponding color factors in the graphical notation are
\begingroup
\allowdisplaybreaks
\begin{align}
  \label{eq:ex_quark_colors}
  \input{tikzExColorsQ},
\end{align}
\endgroup
where the remaining four color factors can be obtained by a permutation of the quark lines $1\leftrightarrow 2$ of the last four color factors in~(\ref{eq:ex_quark_colors}).

The primitive amplitudes in~(\ref{eq:ex_quark_basis}) and the color factors~(\ref{eq:ex_quark_colors}) contain already the complete information for this sector of the amplitudes and can be assembled to obtain~$\mathcal{A}^q_{5,2}$ according to~(\ref{eq:color_dec_quark}).

The basis of primitive amplitudes with a mixed or gluonic loop contains 16 elements. As before we organize the primitive amplitudes line by line according to their quark structure
\begin{equation}
  \label{eq:ex_gluon_basis}
  \begin{aligned}
    \mathcal{B}^g_{5,2} = \big\{
    &A^{g[1^\tL,2^\tL]}(\ul{1}3\ol{1}\ul{2}\ol{2}), A^{g[1^\tL,2^\tL]}(\ul{1}\ol{1}3\ul{2}\ol{2}), A^{g[1^\tL,2^\tL]}(\ul{1}\ol{1}\ul{2}3\ol{2}), A^{g[1^\tL,2^\tL]}(\ul{1}\ol{1}\ul{2}\ol{2}3),\\
    &A^{g[1^\tL,2^\tR]}(\ul{1}3\ol{1}\ol{2}\ul{2}), A^{g[1^\tL,2^\tR]}(\ul{1}\ol{1}3\ol{2}\ul{2}), A^{g[1^\tL,2^\tR]}(\ul{1}\ol{1}\ol{2}3\ul{2}), A^{g[1^\tL,2^\tR]}(\ul{1}\ol{1}\ol{2}\ul{2}3),\\
    &A^{g[1^\tL,2^\tL]}(\ul{1}3\ul{2}\ol{2}\ol{1}), A^{g[1^\tL,2^\tL]}(\ul{1}\ul{2}3\ol{2}\ol{1}), A^{g[1^\tL,2^\tL]}(\ul{1}\ul{2}\ol{2}3\ol{1}), A^{g[1^\tL,2^\tL]}(\ul{1}\ul{2}\ol{2}\ol{1}3),\\
    &A^{g[1^\tL,2^\tL]}(\ul{2}3\ul{1}\ol{1}\ol{2}), A^{g[1^\tL,2^\tL]}(\ul{2}\ul{1}3\ol{1}\ol{2}), A^{g[1^\tL,2^\tL]}(\ul{2}\ul{1}\ol{1}3\ol{2}), A^{g[1^\tL,2^\tL]}(\ul{2}\ul{1}\ol{1}\ol{2}3)\big\}.
  \end{aligned}
\end{equation}
The second line contains the case where the quark line~2 is inverted. This also implies a change of the side on which the gluon loop lies with respect to this line, hence the superscript $2^\tR$. The corresponding color factors according to~(\ref{eq:color_gluon}) have the graphical form
\begingroup
\allowdisplaybreaks
\begin{align}
  \label{eq:ex_gluon_colors}
    \input{tikzExColorsG},
\end{align}
\endgroup
where the remaining four color factors can again be obtained by a permutation of the quark lines $1\leftrightarrow 2$.

The second part of the full amplitude~$\mathcal{A}^g_{5,2}$ is computed from the primitive amplitudes~(\ref{eq:ex_gluon_basis}) and the color factors~(\ref{eq:ex_gluon_colors}) using the decomposition sum in~(\ref{eq:decomp_gluon}). This completes the computation of the full one-loop amplitudes in terms of primitive amplitudes and color factors.

\section{Conclusions and outlook}\label{sec:conclusions}

We have explicitly given a basis of planar primitive one-loop amplitudes sufficient for recovering any color-dressed one-loop QCD amplitude. The purely kinematic primitive amplitudes are gauge-invariant (for theories with no gauge anomalies) and are computed using color-ordered Feynman rules. The number of independent primitive amplitudes for multiplicity~$n$ is in the general case given by $(n-1)!/k!+2^{2k-1}(n-1)!k!/(2k)!$ where $k$ counts the number of quark-antiquark pairs. The first summand is reduced to $(n-2)!(n-k-1)/k!$ for the case of a gauge group with traceless generators. The two contributions are split into amplitudes that contain a closed quark loop and amplitudes that have at least one gluon carrying loop momentum.

The conjectured color decomposition includes a minimal set of linearly independent primitive amplitudes (under all relations with constant coefficients) and color factors independent under the Jacobi identity and the commutation relation. The results are independent of the choice of the gauge group and the number of quark flavors, and are applicable to massless and massive QCD as well as supersymmetric extensions thereof.

High-precision computations of multiple jet events observed at LHC require a high multiplicity of external partons. The new color decomposition has the advantage of being analytically compact and dampens the factorial growth in the number of higher-multiplicity primitive amplitudes. It has as such the potential to improve the efficiency of phenomenological QCD computations.

Compared to results obtained with techniques for a $\text{SU}(N)$ gauge group~\cite{Bern:1994fz,Ita:2011ar,Reuschle:2013qna} we circumvent the step of assembling primitive amplitudes into partial amplitudes and instead provide a direct formula to obtain the full color-dressed amplitude in terms of primitive amplitudes. In the former constructions a given primitive amplitude contributes in general to several partial amplitudes and appears as such several times with a different color factor in the color decomposition. The formulae given here collect all these terms inside one single color factor for every primitive amplitude in the basis. We have checked that our results match the full amplitudes obtained in~\cite{Ita:2011ar,Reuschle:2013qna} at four points.

We provide a Mathematica implementation for the computation of color factors given a (modified) Dyck word and algorithms that produce the basis of primitive amplitudes in terms of Dyck words. The ancillary file is attached to the arXiv submission and contains the examples explicitly worked out in this paper.

For theories with color-kinematics duality~\cite{Bern:2008qj, Bern:2010ue} there exist further relations between primitive amplitudes, called BCJ relations. At tree level a color decomposition using this even smaller subset of independent amplitudes is described in~\cite{Johansson:2015oia}. These relations have also been understood to arise due to a color-factor symmetry~\cite{Brown:2016hck, Brown:2016mrh}. Subsequently we expect the BCJ relations to further reduce of the size of the basis of primitive amplitudes at one-loop~\cite{Boels:2013bi, He:2015wgf, Primo:2016omk, Tourkine:2016bak, He:2016mzd, He:2017spx, Hohenegger:2017kqy, Ochirov:2017jby, Mafra:2017ioj, Geyer:2017ela, Jurado:2017xut}. There is also the hope that a systematic treatment of the color algebra leads to an improved understanding of the kinematic algebra underlying QCD through the color-kinematics duality.

The `Mario world' structure of the results raises the hope of higher-loop generalizations. A remaining challenge is to properly define gauge-invariant primitive amplitudes for non-planar graphs and a basis that induces a compact color decomposition. A simple test case at two-loops is for example~$\mathcal{N}=2$ supersymmetric QCD~\cite{Johansson:2017bfl}.

\begin{acknowledgments}
I would like to thank Henrik Johansson for a lot of enlightening discussions on the topic and feedback on the manuscript. I am also grateful to Konstantina Polydorou for comments on the draft of this paper. The research is supported by the Swedish Research Council under grant 621-2014-5722, the Knut and Alice Wallenberg Foundation under grant KAW 2013.0235, and the Ragnar S\"{o}derberg Foundation under grant S1/16.
\end{acknowledgments}

\appendix

\bibliographystyle{JHEP}
\bibliography{references}

\end{document}

%% file: tikzMarioWorlds.tex
\begin{tikzpicture}
  [>=stealth, baseline=25, inner sep=1]
  \draw[fermion] (1,0) -- (0,0) node[left] {$\ul{1}\strut$};
  \draw[fermion] (2,0) -- (1,0);
  \draw[fermion] (3,0) -- (2,0);
  \draw[fermion] (5,0) -- (3,0);
  \draw[fermion] (6,0) -- (5,0);
  \draw[fermion] (7,0) node[right] {$\ol{1}\strut$} -- (6,0);
   
  \draw[fermion] (1,0.7) -- (0.3,0.7) node[left] {$\ul{2}\strut$};
  \draw[fermion] (2,0.7) -- (1,0.7);
  \draw[fermion] (3,0.7) -- (2,0.7);
  \draw[fermion] (3.7,0.7) node[right] {$\ol{2}\strut$} -- (3,0.7);
  \draw[fermion] (5,0.7) -- (4.3,0.7) node[left] {$\ul{5}\strut$};
  \draw[fermion] (6,0.7) -- (5,0.7);
  \draw[fermion] (6.7,0.7) node[right] {$\ol{5}\strut$} -- (6,0.7);
  \draw[gluon] (1,0) -- (1,0.7);
  \draw[gluon] (2,0) -- (2,0.7);
  \draw[gluon] (3,0) -- (3,0.7);
  \draw[gluon] (5,0) -- (5,0.7);
  \draw[gluon] (6,0) -- (6,0.7);

  \draw[fermion] (2,1.4) -- (1.3,1.4) node[left] {$\ul{3}\strut$};
  \draw[fermion] (3,1.4) -- (2,1.4);
  \draw[fermion] (3.7, 1.4) node[right] {$\ol{3}\strut$} -- (3,1.4);
  \draw[gluon] (2,0.7) -- (2,1.4);
  \draw[gluon] (3,0.7) -- (3,1.4);
  \draw[gluon] (6,0.7) -- (6,1.4) node[above] {6};

  \draw[gluon] (3,1.4) -- (3,2.1) node[above] {4};

  \cC{(1,0)};
  \cC{(2,0)};
  \cC{(3,0)};
  \cC{(5,0)};
  \cC{(6,0)};
  \cC{(2,0.7)};
  \cC{(3,0.7)};
  \cC{(6,0.7)};
  \cC{(3,1.4)};
\end{tikzpicture}

%% file: tikzExClosedQuark.tex
\begin{tikzpicture}
  [>=stealth, baseline=25, inner sep=1]
  \draw[fermion] (1,0) -- (0,0);
  \draw[fermion] (2,0) -- (1,0);
  \draw[fermion] (3,0) -- (2,0);
  \draw[fermion] (5,0) -- (3,0);
  \draw[fermion] (6,0) -- (5,0);
  \draw[fermion] (7,0) -- (6,0);
  \draw[fermion] (0,0) .. controls (-1,-0.5) and (8,-0.5) .. (7,0);
  
  \draw[fermion] (1,0.7) -- (0.3,0.7) node[left] {$\ul{1}\strut$};
  \draw[fermion] (2,0.7) -- (1,0.7);
  \draw[fermion] (3,0.7) -- (2,0.7);
  \draw[fermion] (3.7,0.7) node[right] {$\ol{1}\strut$} -- (3,0.7);
  \draw[fermion] (5,0.7) -- (4.3,0.7) node[left] {$\ul{4}\strut$};
  \draw[fermion] (6,0.7) -- (5,0.7);
  \draw[fermion] (6.7,0.7) node[right] {$\ol{4}\strut$} -- (6,0.7);
  \draw[gluon] (1,0) -- (1,0.7);
  \draw[gluon] (2,0) -- (2,0.7);
  \draw[gluon] (3,0) -- (3,0.7);
  \draw[gluon] (5,0) -- (5,0.7);
  \draw[gluon] (6,0) -- (6,0.7);

  \draw[fermion] (2,1.4) -- (1.3,1.4) node[left] {$\ul{2}\strut$};
  \draw[fermion] (3,1.4) -- (2,1.4);
  \draw[fermion] (3.7, 1.4) node[right] {$\ol{2}\strut$} -- (3,1.4);
  \draw[gluon] (2,0.7) -- (2,1.4);
  \draw[gluon] (3,0.7) -- (3,1.4);
  \draw[gluon] (6,0.7) -- (6,1.4) node[above] {5};

  \draw[gluon] (3,1.4) -- (3,2.1) node[above] {3};

  \cC{(1,0)};
  \cC{(2,0)};
  \cC{(3,0)};
  \cC{(5,0)};
  \cC{(6,0)};
  \cC{(2,0.7)};
  \cC{(3,0.7)};
  \cC{(6,0.7)};
  \cC{(3,1.4)};
\end{tikzpicture}

%% file: tikzExMixed.tex
\begin{tikzpicture}
  [>=stealth, baseline=-0.5, inner sep=1]
  \draw[fermion] (2.5,0) -- (2,0);
  \draw[gluon] (2.5,0) -- (4,0);
  \draw[fermion] (2,0) -- (1.5, 0.7) node[above] {$\ul{1}\strut$};
  \draw[fermion] (3,0.7) node[above] {$\ol{1}\strut$} -- (2.5,0);
  \draw[fermion] (3.5,0.7) node[above] {$\ol{2}\strut$} -- (4,0);
  \draw[fermion] (4,0) -- (5,0);
  \draw[fermion] (5,0) -- (6,0);
  \draw[fermion] (6,0) -- (7,0);
  \draw[fermion] (7,0) -- (7.5,0.7) node[above] {$\ul{2}\strut$};
  \draw[fermion] (4.3,0.7) node[left] {$\ol{3}\strut$} -- (5,0.7);
  \draw[fermion] (5,0.7) -- (6,0.7);
  \draw[fermion] (6,0.7) -- (6.7,0.7) node[right] {$\ul{3}\strut$};
  \draw[gluon] (5,0) -- (5,0.7);
  \draw[gluon] (6,0) -- (6,0.7);
  \draw[gluon] (6,0.7) -- (6,1.4) node[above] {4};
  \draw[gluon] (8.1,0.1) -- (8.1,0.7) node[above] {$5\strut$};
  \draw[gluon] (7,0) -- (8.6,0) .. controls (9.6,-1) and (0,-1) .. (1,0) -- (2,0);

  \cC{(5,0)};
  \cC{(6,0)};
  \cC{(6,0.7)};
\end{tikzpicture}

%% file: tikzExColorsQ.tex
&\begin{tikzpicture}
  [>=stealth, inner sep=1, scale=0.9]
  \draw[fermion] (1,0) -- (0,0);
  \draw[fermion] (2,0) -- (1,0);
  \draw[fermion] (4,0) -- (2,0);
  \draw[fermion] (5,0) -- (4,0);
  \draw[gluon] (1,0) -- (1,0.6);
  \draw[gluon] (2,0) -- (2,0.6);
  \draw[gluon] (4,0) -- (4,0.6);

  \draw[fermion] (1,0.6) -- (0.3,0.6) node[left] {$\ul{1}\strut$};
  \draw[fermion] (2,0.6) -- (1,0.6);
  \draw[fermion] (2.7,0.6) node[right] {$\ol{1}\strut$} -- (2,0.6);
  \draw[gluon] (2,0.6) -- (2,1.2) node[above] {$3$};

  \draw[fermion] (4.7,0.6) node[right] {$\ol{2}\strut$} -- (4,0.6);
  \draw[fermion] (4,0.6) -- (3.3,0.6) node[left] {$\ul{2}\strut$};

  \draw[fermion] (0,0) .. controls (-1,-0.5) and (6,-0.5) .. (5,0);

  \node at (2.5,-0.8) {$C^q(\ul{1}3\ol{1}\ul{2}\ol{2})$};

  \cC{(1,0)};
  \cC{(2,0)};
  \cC{(2,0.6)};
  \cC{(4,0)};
 \end{tikzpicture}
&\begin{tikzpicture}
  [>=stealth, inner sep=1, scale=0.9]
  \draw[fermion] (1,0) -- (0,0);
  \draw[fermion] (2.5,0) -- (1,0);
  \draw[fermion] (4,0) -- (2.5,0);
  \draw[fermion] (5,0) -- (4,0);
  \draw[gluon] (1,0) -- (1,0.6);
  \draw[gluon] (2.5,0) -- (2.5,0.6) node[above] {$3$};
  \draw[gluon] (4,0) -- (4,0.6);

  \draw[fermion] (1,0.6) -- (0.3,0.6) node[left] {$\ul{1}\strut$};
  \draw[fermion] (1.7,0.6) node[right] {$\ol{1}\strut$}-- (1,0.6);

  \draw[fermion] (4.7,0.6) node[right] {$\ol{2}\strut$} -- (4,0.6);
  \draw[fermion] (4,0.6) -- (3.3,0.6) node[left] {$\ul{2}\strut$};

  \draw[fermion] (0,0) .. controls (-1,-0.5) and (6,-0.5) .. (5,0);

  \node at (2.5,-0.8) {$C^q(\ul{1}\ol{1}3\ul{2}\ol{2})$};

  \cC{(1,0)};
  \cC{(2.5,0)};
  \cC{(4,0)};
\end{tikzpicture}\notag\\[-0.5em]
&\begin{tikzpicture}
  [>=stealth, inner sep=1, scale=0.9]
  \draw[fermion] (1,0) -- (0,0);
  \draw[fermion] (3,0) -- (1,0);
  \draw[fermion] (4,0) -- (3,0);
  \draw[fermion] (5,0) -- (4,0);
  \draw[gluon] (1,0) -- (1,0.6);
  \draw[gluon] (3,0) -- (3,0.6);
  \draw[gluon] (4,0) -- (4,0.6);

  \draw[fermion] (1,0.6) -- (0.3,0.6) node[left] {$\ul{1}\strut$};
  \draw[fermion] (1.7,0.6) node[right] {$\ol{1}\strut$}-- (1,0.6);

  \draw[fermion] (4.7,0.6) node[right] {$\ol{2}\strut$} -- (4,0.6);
  \draw[fermion] (4,0.6) -- (3,0.6);
  \draw[fermion] (3,0.6) -- (2.3,0.6) node[left] {$\ul{2}\strut$};
  \draw[gluon] (4,0.6) -- (4,1.2) node[above] {$3$};

  \draw[fermion] (0,0) .. controls (-1,-0.5) and (6,-0.5) .. (5,0);

  \node at (2.5,-0.8) {$C^q(\ul{1}\ol{1}\ul{2}3\ol{2})$};

  \cC{(1,0)};
  \cC{(3,0)};
  \cC{(4,0)};
  \cC{(4,0.6)};
 \end{tikzpicture}
&\begin{tikzpicture}
  [>=stealth, inner sep=1, scale=0.9]
  \draw[fermion] (1,0) -- (0,0);
  \draw[fermion] (3,0) -- (1,0);
  \draw[fermion] (4.5,0) -- (3,0);
  \draw[fermion] (5,0) -- (4.5,0);
  \draw[gluon] (1,0) -- (1,0.6);
  \draw[gluon] (3,0) -- (3,0.6);
  \draw[fermion] (0,0) .. controls (-1,-0.5) and (6,-0.5) .. (5,0);

  \draw[fermion] (1,0.6) -- (0.3,0.6) node[left] {$\ul{1}\strut$};
  \draw[fermion] (1.7,0.6) node[right] {$\ol{1}\strut$}-- (1,0.6);

  \draw[fermion] (3.7,0.6) node[right] {$\ol{2}\strut$} -- (3,0.6);
  \draw[fermion] (3,0.6) -- (2.3,0.6) node[left] {$\ul{2}\strut$};

  \draw[gluon] (4.5,0) -- (4.5,0.6) node[above] {$3$};

  \node at (2.5,-0.8) {$C^q(\ul{1}\ol{1}\ul{2}\ol{2}3)$};

  \cC{(1,0)};
  \cC{(3,0)};
  \cC{(4.5,0)};
 \end{tikzpicture}\\[-0.5em]
&\begin{tikzpicture}
  [>=stealth, inner sep=1, scale=0.9]
  \draw[fermion] (1,0) -- (0,0);
  \draw[fermion] (2.5,0) -- (1,0);
  \draw[fermion] (4,0) -- (2.5,0);
  \draw[fermion] (5,0) -- (4,0);
  \draw[gluon] (1,0) -- (1,0.6);
  \draw[gluon] (2.5,0) -- (2.5,0.6);
  \draw[gluon] (4,0) -- (4,0.6);

  \draw[fermion] (1,0.6) -- (0.3,0.6) node[left] {$\ul{1}\strut$};
  \draw[fermion] (2.5,0.6) -- (1,0.6);
  \draw[fermion] (4,0.6) -- (2.5,0.6);
  \draw[fermion] (4.7,0.6) node[right] {$\ol{1}\strut$} -- (4,0.6);
  \draw[gluon] (2.5,0.6) -- (2.5,1.2) node[above] {$3$};
  \draw[gluon] (4,0.6) -- (4,1.2);

  \draw[fermion] (4.7,1.2) node[right] {$\ol{2}\strut$} -- (4,1.2);
  \draw[fermion] (4,1.2) -- (3.3,1.2) node[left] {$\ul{2}\strut$};

  \draw[fermion] (0,0) .. controls (-1,-0.5) and (6,-0.5) .. (5,0);

  \node at (2.5,-0.8) {$C^q(\ul{1}3\ul{2}\ol{2}\ol{1})$};

  \cC{(1,0)};
  \cC{(2.5,0)};
  \cC{(2.5,0.6)};
  \cC{(4,0)};
  \cC{(4,0.6)};
 \end{tikzpicture}
&\begin{tikzpicture}
  [>=stealth, inner sep=1, scale=0.9]
  \draw[fermion] (1,0) -- (0,0);
  \draw[fermion] (2.5,0) -- (1,0);
  \draw[fermion] (4,0) -- (2.5,0);
  \draw[fermion] (5,0) -- (4,0);
  \draw[gluon] (1,0) -- (1,0.6);
  \draw[gluon] (2.5,0) -- (2.5,0.6);
  \draw[gluon] (4,0) -- (4,0.6);

  \draw[fermion] (1,0.6) -- (0.3,0.6) node[left] {$\ul{1}\strut$};
  \draw[fermion] (2.5,0.6) -- (1,0.6);
  \draw[fermion] (4,0.6) -- (2.5,0.6);
  \draw[fermion] (4.7,0.6) node[right] {$\ol{1}\strut$} -- (4,0.6);
  \draw[gluon] (2.5,0.6) -- (2.5,1.2);
  \draw[gluon] (4,0.6) -- (4,1.2);

  \draw[fermion] (4.7,1.2) node[right] {$\ol{2}\strut$} -- (4,1.2);
  \draw[fermion] (4,1.2) -- (2.5,1.2);
  \draw[fermion] (2.5,1.2) -- (1.8,1.2) node[left] {$\ul{2}\strut$};
  \draw[gluon] (4,1.2) -- (4,1.8) node[above] {$3$};

  \draw[fermion] (0,0) .. controls (-1,-0.5) and (6,-0.5) .. (5,0);

  \node at (2.5,-0.8) {$C^q(\ul{1}\ul{2}3\ol{2}\ol{1})$};

  \cC{(1,0)};
  \cC{(2.5,0)};
  \cC{(2.5,0.6)};
  \cC{(4,0)};
  \cC{(4,0.6)};
  \cC{(4,1.2)};
 \end{tikzpicture}\notag\\[-0.5em]
&\begin{tikzpicture}
  [>=stealth, inner sep=1, scale=0.9]
  \draw[fermion] (1,0) -- (0,0);
  \draw[fermion] (2.5,0) -- (1,0);
  \draw[fermion] (4,0) -- (2.5,0);
  \draw[fermion] (5,0) -- (4,0);
  \draw[gluon] (1,0) -- (1,0.6);
  \draw[gluon] (2.5,0) -- (2.5,0.6);
  \draw[gluon] (4,0) -- (4,0.6);

  \draw[fermion] (1,0.6) -- (0.3,0.6) node[left] {$\ul{1}\strut$};
  \draw[fermion] (2.5,0.6) -- (1,0.6);
  \draw[fermion] (4,0.6) -- (2.5,0.6);
  \draw[fermion] (4.7,0.6) node[right] {$\ol{1}\strut$} -- (4,0.6);
  \draw[gluon] (2.5,0.6) -- (2.5,1.2);
  \draw[gluon] (4,0.6) -- (4,1.2) node[above] {$3$};

  \draw[fermion] (3.2,1.2) node[right] {$\ol{2}\strut$} -- (2.5,1.2);
  \draw[fermion] (2.5,1.2) -- (1.8,1.2) node[left] {$\ul{2}\strut$};

  \draw[fermion] (0,0) .. controls (-1,-0.5) and (6,-0.5) .. (5,0);

  \node at (2.5,-0.8) {$C^q(\ul{1}\ul{2}\ol{2}3\ol{1})$};

  \cC{(1,0)};
  \cC{(2.5,0)};
  \cC{(2.5,0.6)};
  \cC{(4,0)};
  \cC{(4,0.6)};
 \end{tikzpicture}
&\begin{tikzpicture}
  [>=stealth, inner sep=1, scale=0.9]
  \draw[fermion] (1,0) -- (0,0);
  \draw[fermion] (2.5,0) -- (1,0);
  \draw[fermion] (4,0) -- (2.5,0);
  \draw[fermion] (5,0) -- (4,0);
  \draw[gluon] (1,0) -- (1,0.6);
  \draw[gluon] (2.5,0) -- (2.5,0.6);
  \draw[gluon] (4,0) -- (4,0.6) node[above] {$3$};

  \draw[fermion] (1,0.6) -- (0.3,0.6) node[left] {$\ul{1}\strut$};
  \draw[fermion] (2.5,0.6) -- (1,0.6);
  \draw[fermion] (3.2,0.6) node[right] {$\ol{1}\strut$} -- (2.5,0.6);
  \draw[gluon] (2.5,0.6) -- (2.5,1.2);

  \draw[fermion] (3.2,1.2) node[right] {$\ol{2}\strut$} -- (2.5,1.2);
  \draw[fermion] (2.5,1.2) -- (1.8,1.2) node[left] {$\ul{2}\strut$};

  \draw[fermion] (0,0) .. controls (-1,-0.5) and (6,-0.5) .. (5,0);

  \node at (2.5,-0.8) {$C^q(\ul{1}\ul{2}\ol{2}\ol{1}3)$};

  \cC{(1,0)};
  \cC{(2.5,0)};
  \cC{(2.5,0.6)};
  \cC{(4,0)};
 \end{tikzpicture}\notag

%% file: tikzExColorsG.tex
&\begin{tikzpicture}
  [>=stealth, inner sep=1, scale=0.9]
  \draw[gluon] (4,0) -- (5,0) .. controls (6,-0.6) and (-1,-0.6) .. (0,0) -- (1,0);
  \draw[fermion] (1,0) -- (0.5,0.6) node[above=-0.05] {$\ul{1}\strut$};
  \draw[fermion] (1.5,0) -- (1,0);
  \draw[fermion] (2,0) -- (1.5,0);
  \draw[fermion] (2.5,0.6) node[above=-0.05] {$\ol{1}\strut$} -- (2,0);
  \draw[gluon] (2,0) -- (3.5,0);
  \draw[fermion] (3.5,0) -- (3,0.6) node[above=-0.05] {$\ul{2}\strut$};
  \draw[fermion] (4,0) -- (3.5,0);
  \draw[fermion] (4.5,0.6) node[above=-0.05] {$\ol{2}\strut$} -- (4,0);

  \draw[gluon] (1.5,0) -- (1.5,0.6) node[above=-0.05] {$3\strut$};

  \node at (2.25,-0.9) {$C^g(\ul{1}3\ol{1}\ul{2}\ol{2})$};
  
  \cC{(1.5,0)};
 \end{tikzpicture}
&\begin{tikzpicture}
  [>=stealth, inner sep=1, scale=0.9]
  \draw[gluon] (4,0) -- (5,0) .. controls (6,-0.6) and (-1,-0.6) .. (0,0) -- (1,0);
  \draw[fermion] (1,0) -- (0.5,0.6) node[above=-0.05] {$\ul{1}\strut$};
  \draw[fermion] (1.5,0) -- (1,0);
  \draw[fermion] (2,0.6) node[above=-0.05] {$\ol{1}\strut$} -- (1.5,0);
  \draw[gluon] (1.5,0) -- (3.5,0);
  \draw[fermion] (3.5,0) -- (3,0.6) node[above=-0.05] {$\ul{2}\strut$};
  \draw[fermion] (4,0) -- (3.5,0);
  \draw[fermion] (4.5,0.6) node[above=-0.05] {$\ol{2}\strut$} -- (4,0);

  \draw[gluon] (2.5,0.1) -- (2.5,0.6) node[above=-0.05] {$3\strut$};

  \node at (2.25,-0.9) {$C^g(\ul{1}\ol{1}3\ul{2}\ol{2})$};
 \end{tikzpicture}\notag\\[-0.5em]
&\begin{tikzpicture}
  [>=stealth, inner sep=1, scale=0.9]
  \draw[gluon] (4,0) -- (5,0) .. controls (6,-0.6) and (-1,-0.6) .. (0,0) -- (1,0);
  \draw[fermion] (1,0) -- (0.5,0.6) node[above=-0.05] {$\ul{1}\strut$};
  \draw[fermion] (1.5,0) -- (1,0);
  \draw[fermion] (2,0.6) node[above=-0.05] {$\ol{1}\strut$} -- (1.5,0);
  \draw[gluon] (1.5,0) -- (3,0);
  \draw[fermion] (3,0) -- (2.5,0.6) node[above=-0.05] {$\ul{2}\strut$};
  \draw[fermion] (3.5,0) -- (3,0);
  \draw[fermion] (4,0) -- (3.5,0);
  \draw[fermion] (4.5,0.6) node[above=-0.05] {$\ol{2}\strut$} -- (4,0);

  \draw[gluon] (3.5,0) -- (3.5,0.6) node[above=-0.05] {$3\strut$};

  \node at (2.25,-0.9) {$C^g(\ul{1}\ol{1}\ul{2}3\ol{2})$};

  \cC{(3.5,0)};
\end{tikzpicture}
&\begin{tikzpicture}
  [>=stealth, inner sep=1, scale=0.9]
  \draw[gluon] (3.5,0) -- (5,0) .. controls (6,-0.6) and (-1,-0.6) .. (0,0) -- (1,0);
  \draw[fermion] (1,0) -- (0.5,0.6) node[above=-0.05] {$\ul{1}\strut$};
  \draw[fermion] (1.5,0) -- (1,0);
  \draw[fermion] (2,0.6) node[above=-0.05] {$\ol{1}\strut$} -- (1.5,0);
  \draw[gluon] (1.5,0) -- (3,0);
  \draw[fermion] (3,0) -- (2.5,0.6) node[above=-0.05] {$\ul{2}\strut$};
  \draw[fermion] (3.5,0) -- (3,0);
  \draw[fermion] (4,0.6) node[above=-0.05] {$\ol{2}\strut$} -- (3.5,0);

  \draw[gluon] (4.5,0.1) -- (4.5,0.6) node[above=-0.05] {$3\strut$};

  \node at (2.25,-0.9) {$C^g(\ul{1}\ol{1}\ul{2}\ol{2}3)$};
 \end{tikzpicture}\notag\\[-0.5em]
&\begin{tikzpicture}
  [>=stealth, inner sep=1, scale=0.9]
  \draw[gluon] (4,0) -- (5,0) .. controls (6,-0.6) and (-1,-0.6) .. (0,0) -- (1,0);
  \draw[fermion] (1,0) -- (0.5,0.6) node[above=-0.05] {$\ul{1}\strut$};
  \draw[fermion] (1.5,0) -- (1,0);
  \draw[fermion] (2,0) -- (1.5,0);
  \draw[fermion] (2.5,0.6) node[above=-0.05] {$\ol{1}\strut$} -- (2,0);
  \draw[gluon] (2,0) -- (3.5,0);
  \draw[fermion] (3,0.6) node[above=-0.05] {$\ol{2}\strut$} -- (3.5,0);
  \draw[fermion] (3.5,0) -- (4,0);
  \draw[fermion] (4,0) -- (4.5,0.6) node[above=-0.05] {$\ul{2}\strut$};

  \draw[gluon] (1.5,0) -- (1.5,0.6) node[above=-0.05] {$3\strut$};

  \node at (2.25,-0.9) {$C^g(\ul{1}3\ol{1}\ol{2}\ul{2})$};
  
  \cC{(1.5,0)};
 \end{tikzpicture}
&\begin{tikzpicture}
  [>=stealth, inner sep=1, scale=0.9]
  \draw[gluon] (4,0) -- (5,0) .. controls (6,-0.6) and (-1,-0.6) .. (0,0) -- (1,0);
  \draw[fermion] (1,0) -- (0.5,0.6) node[above=-0.05] {$\ul{1}\strut$};
  \draw[fermion] (1.5,0) -- (1,0);
  \draw[fermion] (2,0.6) node[above=-0.05] {$\ol{1}\strut$} -- (1.5,0);
  \draw[gluon] (1.5,0) -- (3.5,0);
  \draw[fermion] (3,0.6) node[above=-0.05] {$\ol{2}\strut$} -- (3.5,0);
  \draw[fermion] (3.5,0) -- (4,0);
  \draw[fermion] (4,0) -- (4.5,0.6) node[above=-0.05] {$\ul{2}\strut$};

  \draw[gluon] (2.5,0.1) -- (2.5,0.6) node[above=-0.05] {$3\strut$};

  \node at (2.25,-0.9) {$C^g(\ul{1}\ol{1}3\ol{2}\ul{2})$};
 \end{tikzpicture}\\[-0.5em]
&\begin{tikzpicture}
  [>=stealth, inner sep=1, scale=0.9]
  \draw[gluon] (4,0) -- (5,0) .. controls (6,-0.6) and (-1,-0.6) .. (0,0) -- (1,0);
  \draw[fermion] (1,0) -- (0.5,0.6) node[above=-0.05] {$\ul{1}\strut$};
  \draw[fermion] (1.5,0) -- (1,0);
  \draw[fermion] (2,0.6) node[above=-0.05] {$\ol{1}\strut$} -- (1.5,0);
  \draw[gluon] (1.5,0) -- (3,0);
  \draw[fermion] (2.5,0.6) node[above=-0.05] {$\ol{2}\strut$} -- (3,0);
  \draw[fermion] (3,0) -- (3.5,0);
  \draw[fermion] (3.5,0) -- (4,0);
  \draw[fermion] (4,0) -- (4.5,0.6) node[above=-0.05] {$\ul{2}\strut$};

  \draw[gluon] (3.5,0) -- (3.5,0.6) node[above=-0.05] {$3\strut$};

  \node at (2.25,-0.9) {$C^g(\ul{1}\ol{1}\ol{2}3\ul{2})$};

  \cC{(3.5,0)};
\end{tikzpicture}
&\begin{tikzpicture}
  [>=stealth, inner sep=1, scale=0.9]
  \draw[gluon] (3.5,0) -- (5,0) .. controls (6,-0.6) and (-1,-0.6) .. (0,0) -- (1,0);
  \draw[fermion] (1,0) -- (0.5,0.6) node[above=-0.05] {$\ul{1}\strut$};
  \draw[fermion] (1.5,0) -- (1,0);
  \draw[fermion] (2,0.6) node[above=-0.05] {$\ol{1}\strut$} -- (1.5,0);
  \draw[gluon] (1.5,0) -- (3,0);
  \draw[fermion] (2.5,0.6) node[above=-0.05] {$\ol{2}\strut$} -- (3,0);
  \draw[fermion] (3,0) -- (3.5,0);
  \draw[fermion] (3.5,0) -- (4,0.6) node[above=-0.05] {$\ul{2}\strut$};

  \draw[gluon] (4.5,0.1) -- (4.5,0.6) node[above=-0.05] {$3\strut$};

  \node at (2.25,-0.9) {$C^g(\ul{1}\ol{1}\ol{2}\ul{2}3)$};
 \end{tikzpicture}\notag\\[-0.5em]
&\begin{tikzpicture}
  [>=stealth, inner sep=1, scale=0.9]
  \draw[gluon] (4,0) -- (5,0) .. controls (6,-0.6) and (-1,-0.6) .. (0,0) -- (1,0);
  \draw[fermion] (1,0) -- (0.5,0.6) node[above=-0.05] {$\ul{1}\strut$};
  \draw[fermion] (1.5,0) -- (1,0);
  \draw[fermion] (3,0) -- (1.5,0);
  \draw[fermion] (4,0) -- (3,0);
  \draw[fermion] (4.5,0.6) node[above=-0.05] {$\ol{1}\strut$} -- (4,0);

  \draw[gluon] (1.5,0) -- (1.5,0.6) node[above=-0.05] {$3\strut$};
  \draw[gluon] (3,0) -- (3,0.6);
  \draw[fermion] (3,0.6) -- (2.3,0.6) node[left] {$\ul{2}\strut$};
  \draw[fermion] (3.7,0.6) node[right] {$\ol{2}\strut$} -- (3,0.6);

  \node at (2.25,-0.9) {$C^g(\ul{1}3\ul{2}\ol{2}\ol{1})$};

  \cC{(1.5,0)};
  \cC{(3,0)};
 \end{tikzpicture}
&\begin{tikzpicture}
  [>=stealth, inner sep=1, scale=0.9]
  \draw[gluon] (4,0) -- (5,0) .. controls (6,-0.6) and (-1,-0.6) .. (0,0) -- (1,0);
  \draw[fermion] (1,0) -- (0.5,0.6) node[above=-0.05] {$\ul{1}\strut$};
  \draw[fermion] (2,0) -- (1,0);
  \draw[fermion] (3,0) -- (2,0);
  \draw[fermion] (4,0) -- (3,0);
  \draw[fermion] (4.5,0.6) node[above=-0.05] {$\ol{1}\strut$} -- (4,0);

  \draw[gluon] (2,0) -- (2,0.6);
  \draw[gluon] (3,0) -- (3,0.6);
  \draw[fermion] (2,0.6) -- (1.3,0.6) node[left] {$\ul{2}\strut$};
  \draw[fermion] (3,0.6) -- (2,0.6);
  \draw[fermion] (3.7,0.6) node[right] {$\ol{2}\strut$} -- (3,0.6);

  \draw[gluon] (3,0.6) -- (3,1.2) node[above=-0.05] {$3\strut$};

  \node at (2.25,-0.9) {$C^g(\ul{1}\ul{2}3\ol{2}\ol{1})$};
  
  \cC{(2,0)};
  \cC{(3,0)};
  \cC{(3,0.6)};
 \end{tikzpicture}\notag\\[-0.5em]
&\begin{tikzpicture}
  [>=stealth, inner sep=1, scale=0.9]
  \draw[gluon] (4,0) -- (5,0) .. controls (6,-0.6) and (-1,-0.6) .. (0,0) -- (1,0);
  \draw[fermion] (1,0) -- (0.5,0.6) node[above=-0.05] {$\ul{1}\strut$};
  \draw[fermion] (2,0) -- (1,0);
  \draw[fermion] (3.5,0) -- (2,0);
  \draw[fermion] (4,0) -- (3.5,0);
  \draw[fermion] (4.5,0.6) node[above=-0.05] {$\ol{1}\strut$} -- (4,0);

  \draw[gluon] (2,0) -- (2,0.6);
  \draw[gluon] (3.5,0) -- (3.5,0.6) node[above=-0.05] {$3\strut$};
  \draw[fermion] (2,0.6) -- (1.3,0.6) node[left] {$\ul{2}\strut$};
  \draw[fermion] (2.7,0.6) node[right] {$\ol{2}\strut$} -- (2,0.6);

  \node at (2.25,-0.9) {$C^g(\ul{1}\ul{2}\ol{2}3\ol{1})$};
  
  \cC{(2,0)};
  \cC{(3.5,0)};
 \end{tikzpicture}
&\begin{tikzpicture}
  [>=stealth, inner sep=1, scale=0.9]
  \draw[gluon] (3,0) -- (5,0) .. controls (6,-0.6) and (-1,-0.6) .. (0,0) -- (1,0);
  \draw[fermion] (1,0) -- (0.5,0.6) node[above=-0.05] {$\ul{1}\strut$};
  \draw[fermion] (2,0) -- (1,0);
  \draw[fermion] (3,0) -- (2,0);
  \draw[fermion] (3.5,0.6) node[above=-0.05] {$\ol{1}\strut$} -- (3,0);

  \draw[gluon] (2,0) -- (2,0.6);  
  \draw[fermion] (2,0.6) -- (1.3,0.6) node[left] {$\ul{2}\strut$};
  \draw[fermion] (2.7,0.6) node[right] {$\ol{2}\strut$} -- (2,0.6);

  \draw[gluon] (4,0.1) -- (4,0.6) node[above=-0.05] {$3\strut$};

  \node at (2.25,-0.9) {$C^g(\ul{1}\ul{2}\ol{2}\ol{1}3)$};
  
  \cC{(2,0)};
\end{tikzpicture}\notag

%% file: oneLoopColor.bbl
\providecommand{\href}[2]{#2}\begingroup\raggedright\begin{thebibliography}{100}

\bibitem{Bern:1990cu}
Z.~Bern and D.~A. Kosower, \emph{{Efficient calculation of one loop QCD
  amplitudes}},
  \href{http://dx.doi.org/10.1103/PhysRevLett.66.1669}{\emph{Phys. Rev. Lett.}
  {\bf 66} (1991) 1669--1672}.

\bibitem{Britto:2004nc}
R.~Britto, F.~Cachazo and B.~Feng, \emph{{Generalized unitarity and one-loop
  amplitudes in N=4 super-Yang-Mills}},
  \href{http://dx.doi.org/10.1016/j.nuclphysb.2005.07.014}{\emph{Nucl. Phys.}
  {\bf B725} (2005) 275--305}, [\href{http://arxiv.org/abs/hep-th/0412103}{{\tt
  hep-th/0412103}}].

\bibitem{Forde:2007mi}
D.~Forde, \emph{{Direct extraction of one-loop integral coefficients}},
  \href{http://dx.doi.org/10.1103/PhysRevD.75.125019}{\emph{Phys. Rev.} {\bf
  D75} (2007) 125019}, [\href{http://arxiv.org/abs/0704.1835}{{\tt
  0704.1835}}].

\bibitem{Ossola:2006us}
G.~Ossola, C.~G. Papadopoulos and R.~Pittau, \emph{{Reducing full one-loop
  amplitudes to scalar integrals at the integrand level}},
  \href{http://dx.doi.org/10.1016/j.nuclphysb.2006.11.012}{\emph{Nucl. Phys.}
  {\bf B763} (2007) 147--169}, [\href{http://arxiv.org/abs/hep-ph/0609007}{{\tt
  hep-ph/0609007}}].

\bibitem{Anastasiou:2006jv}
C.~Anastasiou, R.~Britto, B.~Feng, Z.~Kunszt and P.~Mastrolia,
  \emph{{D-dimensional unitarity cut method}},
  \href{http://dx.doi.org/10.1016/j.physletb.2006.12.022}{\emph{Phys. Lett.}
  {\bf B645} (2007) 213--216}, [\href{http://arxiv.org/abs/hep-ph/0609191}{{\tt
  hep-ph/0609191}}].

\bibitem{Anastasiou:2006gt}
C.~Anastasiou, R.~Britto, B.~Feng, Z.~Kunszt and P.~Mastrolia, \emph{{Unitarity
  cuts and Reduction to master integrals in d dimensions for one-loop
  amplitudes}},
  \href{http://dx.doi.org/10.1088/1126-6708/2007/03/111}{\emph{JHEP} {\bf 03}
  (2007) 111}, [\href{http://arxiv.org/abs/hep-ph/0612277}{{\tt
  hep-ph/0612277}}].

\bibitem{Giele:2008ve}
W.~T. Giele, Z.~Kunszt and K.~Melnikov, \emph{{Full one-loop amplitudes from
  tree amplitudes}},
  \href{http://dx.doi.org/10.1088/1126-6708/2008/04/049}{\emph{JHEP} {\bf 04}
  (2008) 049}, [\href{http://arxiv.org/abs/0801.2237}{{\tt 0801.2237}}].

\bibitem{Giele:2008bc}
W.~T. Giele and G.~Zanderighi, \emph{{On the Numerical Evaluation of One-Loop
  Amplitudes: The Gluonic Case}},
  \href{http://dx.doi.org/10.1088/1126-6708/2008/06/038}{\emph{JHEP} {\bf 06}
  (2008) 038}, [\href{http://arxiv.org/abs/0805.2152}{{\tt 0805.2152}}].

\bibitem{Ellis:2008ir}
R.~K. Ellis, W.~T. Giele, Z.~Kunszt and K.~Melnikov, \emph{{Masses, fermions
  and generalized $D$-dimensional unitarity}},
  \href{http://dx.doi.org/10.1016/j.nuclphysb.2009.07.023}{\emph{Nucl. Phys.}
  {\bf B822} (2009) 270--282}, [\href{http://arxiv.org/abs/0806.3467}{{\tt
  0806.3467}}].

\bibitem{Ellis:2008qc}
R.~K. Ellis, W.~T. Giele, Z.~Kunszt, K.~Melnikov and G.~Zanderighi,
  \emph{{One-loop amplitudes for $W^+$ 3 jet production in hadron collisions}},
  \href{http://dx.doi.org/10.1088/1126-6708/2009/01/012}{\emph{JHEP} {\bf 01}
  (2009) 012}, [\href{http://arxiv.org/abs/0810.2762}{{\tt 0810.2762}}].

\bibitem{Assadsolimani:2009cz}
M.~Assadsolimani, S.~Becker and S.~Weinzierl, \emph{{A Simple formula for the
  infrared singular part of the integrand of one-loop QCD amplitudes}},
  \href{http://dx.doi.org/10.1103/PhysRevD.81.094002}{\emph{Phys. Rev.} {\bf
  D81} (2010) 094002}, [\href{http://arxiv.org/abs/0912.1680}{{\tt
  0912.1680}}].

\bibitem{Becker:2010ng}
S.~Becker, C.~Reuschle and S.~Weinzierl, \emph{{Numerical NLO QCD
  calculations}}, \href{http://dx.doi.org/10.1007/JHEP12(2010)013}{\emph{JHEP}
  {\bf 12} (2010) 013}, [\href{http://arxiv.org/abs/1010.4187}{{\tt
  1010.4187}}].

\bibitem{Becker:2012aqa}
S.~Becker, C.~Reuschle and S.~Weinzierl, \emph{{Efficiency Improvements for the
  Numerical Computation of NLO Corrections}},
  \href{http://dx.doi.org/10.1007/JHEP07(2012)090}{\emph{JHEP} {\bf 07} (2012)
  090}, [\href{http://arxiv.org/abs/1205.2096}{{\tt 1205.2096}}].

\bibitem{Ellis:2011cr}
R.~K. Ellis, Z.~Kunszt, K.~Melnikov and G.~Zanderighi, \emph{{One-loop
  calculations in quantum field theory: from Feynman diagrams to unitarity
  cuts}}, \href{http://dx.doi.org/10.1016/j.physrep.2012.01.008}{\emph{Phys.
  Rept.} {\bf 518} (2012) 141--250},
  [\href{http://arxiv.org/abs/1105.4319}{{\tt 1105.4319}}].

\bibitem{Bern:2013gka}
Z.~Bern, L.~J. Dixon, F.~Febres~Cordero, S.~Höche, H.~Ita, D.~A. Kosower
  et~al., \emph{{Next-to-Leading Order $W + 5$-Jet Production at the LHC}},
  \href{http://dx.doi.org/10.1103/PhysRevD.88.014025}{\emph{Phys. Rev.} {\bf
  D88} (2013) 014025}, [\href{http://arxiv.org/abs/1304.1253}{{\tt
  1304.1253}}].

\bibitem{Bonciani:2015eua}
R.~Bonciani, V.~Del~Duca, H.~Frellesvig, J.~M. Henn, F.~Moriello and V.~A.
  Smirnov, \emph{{Next-to-leading order QCD corrections to the decay width H
  $\rightarrow$ Z$\gamma$}},
  \href{http://dx.doi.org/10.1007/JHEP08(2015)108}{\emph{JHEP} {\bf 08} (2015)
  108}, [\href{http://arxiv.org/abs/1505.00567}{{\tt 1505.00567}}].

\bibitem{Bern:1994zx}
Z.~Bern, L.~J. Dixon, D.~C. Dunbar and D.~A. Kosower, \emph{{One loop n point
  gauge theory amplitudes, unitarity and collinear limits}},
  \href{http://dx.doi.org/10.1016/0550-3213(94)90179-1}{\emph{Nucl. Phys.} {\bf
  B425} (1994) 217--260}, [\href{http://arxiv.org/abs/hep-ph/9403226}{{\tt
  hep-ph/9403226}}].

\bibitem{Bern:1994cg}
Z.~Bern, L.~J. Dixon, D.~C. Dunbar and D.~A. Kosower, \emph{{Fusing gauge
  theory tree amplitudes into loop amplitudes}},
  \href{http://dx.doi.org/10.1016/0550-3213(94)00488-Z}{\emph{Nucl. Phys.} {\bf
  B435} (1995) 59--101}, [\href{http://arxiv.org/abs/hep-ph/9409265}{{\tt
  hep-ph/9409265}}].

\bibitem{Bern:1996je}
Z.~Bern, L.~J. Dixon and D.~A. Kosower, \emph{{Progress in one loop QCD
  computations}},
  \href{http://dx.doi.org/10.1146/annurev.nucl.46.1.109}{\emph{Ann. Rev. Nucl.
  Part. Sci.} {\bf 46} (1996) 109--148},
  [\href{http://arxiv.org/abs/hep-ph/9602280}{{\tt hep-ph/9602280}}].

\bibitem{Ellis:2007br}
R.~K. Ellis, W.~T. Giele and Z.~Kunszt, \emph{{A Numerical Unitarity Formalism
  for Evaluating One-Loop Amplitudes}},
  \href{http://dx.doi.org/10.1088/1126-6708/2008/03/003}{\emph{JHEP} {\bf 03}
  (2008) 003}, [\href{http://arxiv.org/abs/0708.2398}{{\tt 0708.2398}}].

\bibitem{Berger:2008sj}
C.~F. Berger, Z.~Bern, L.~J. Dixon, F.~Febres~Cordero, D.~Forde, H.~Ita et~al.,
  \emph{{An Automated Implementation of On-Shell Methods for One-Loop
  Amplitudes}}, \href{http://dx.doi.org/10.1103/PhysRevD.78.036003}{\emph{Phys.
  Rev.} {\bf D78} (2008) 036003}, [\href{http://arxiv.org/abs/0803.4180}{{\tt
  0803.4180}}].

\bibitem{Ossola:2007ax}
G.~Ossola, C.~G. Papadopoulos and R.~Pittau, \emph{{CutTools: A Program
  implementing the OPP reduction method to compute one-loop amplitudes}},
  \href{http://dx.doi.org/10.1088/1126-6708/2008/03/042}{\emph{JHEP} {\bf 03}
  (2008) 042}, [\href{http://arxiv.org/abs/0711.3596}{{\tt 0711.3596}}].

\bibitem{Mastrolia:2008jb}
P.~Mastrolia, G.~Ossola, C.~G. Papadopoulos and R.~Pittau, \emph{{Optimizing
  the Reduction of One-Loop Amplitudes}},
  \href{http://dx.doi.org/10.1088/1126-6708/2008/06/030}{\emph{JHEP} {\bf 06}
  (2008) 030}, [\href{http://arxiv.org/abs/0803.3964}{{\tt 0803.3964}}].

\bibitem{Mastrolia:2010nb}
P.~Mastrolia, G.~Ossola, T.~Reiter and F.~Tramontano, \emph{{Scattering
  Amplitudes from Unitarity-based Reduction Algorithm at the Integrand-level}},
  \href{http://dx.doi.org/10.1007/JHEP08(2010)080}{\emph{JHEP} {\bf 08} (2010)
  080}, [\href{http://arxiv.org/abs/1006.0710}{{\tt 1006.0710}}].

\bibitem{Badger:2010nx}
S.~Badger, B.~Biedermann and P.~Uwer, \emph{{NGluon: A Package to Calculate
  One-loop Multi-gluon Amplitudes}},
  \href{http://dx.doi.org/10.1016/j.cpc.2011.04.008}{\emph{Comput. Phys.
  Commun.} {\bf 182} (2011) 1674--1692},
  [\href{http://arxiv.org/abs/1011.2900}{{\tt 1011.2900}}].

\bibitem{Hirschi:2011pa}
V.~Hirschi, R.~Frederix, S.~Frixione, M.~V. Garzelli, F.~Maltoni and R.~Pittau,
  \emph{{Automation of one-loop QCD corrections}},
  \href{http://dx.doi.org/10.1007/JHEP05(2011)044}{\emph{JHEP} {\bf 05} (2011)
  044}, [\href{http://arxiv.org/abs/1103.0621}{{\tt 1103.0621}}].

\bibitem{Berends:1987me}
F.~A. Berends and W.~T. Giele, \emph{{Recursive Calculations for Processes with
  n Gluons}}, \href{http://dx.doi.org/10.1016/0550-3213(88)90442-7}{\emph{Nucl.
  Phys.} {\bf B306} (1988) 759--808}.

\bibitem{Britto:2004ap}
R.~Britto, F.~Cachazo and B.~Feng, \emph{{New recursion relations for tree
  amplitudes of gluons}},
  \href{http://dx.doi.org/10.1016/j.nuclphysb.2005.02.030}{\emph{Nucl. Phys.}
  {\bf B715} (2005) 499--522}, [\href{http://arxiv.org/abs/hep-th/0412308}{{\tt
  hep-th/0412308}}].

\bibitem{Britto:2005fq}
R.~Britto, F.~Cachazo, B.~Feng and E.~Witten, \emph{{Direct proof of tree-level
  recursion relation in Yang-Mills theory}},
  \href{http://dx.doi.org/10.1103/PhysRevLett.94.181602}{\emph{Phys. Rev.
  Lett.} {\bf 94} (2005) 181602},
  [\href{http://arxiv.org/abs/hep-th/0501052}{{\tt hep-th/0501052}}].

\bibitem{Mafra:2010jq}
C.~R. Mafra, O.~Schlotterer, S.~Stieberger and D.~Tsimpis, \emph{{A recursive
  method for SYM n-point tree amplitudes}},
  \href{http://dx.doi.org/10.1103/PhysRevD.83.126012}{\emph{Phys. Rev.} {\bf
  D83} (2011) 126012}, [\href{http://arxiv.org/abs/1012.3981}{{\tt
  1012.3981}}].

\bibitem{Cachazo:2004kj}
F.~Cachazo, P.~Svrcek and E.~Witten, \emph{{MHV vertices and tree amplitudes in
  gauge theory}},
  \href{http://dx.doi.org/10.1088/1126-6708/2004/09/006}{\emph{JHEP} {\bf 09}
  (2004) 006}, [\href{http://arxiv.org/abs/hep-th/0403047}{{\tt
  hep-th/0403047}}].

\bibitem{vanOldenborgh:1989wn}
G.~J. van Oldenborgh and J.~A.~M. Vermaseren, \emph{{New Algorithms for One
  Loop Integrals}}, \href{http://dx.doi.org/10.1007/BF01621031}{\emph{Z. Phys.}
  {\bf C46} (1990) 425--438}.

\bibitem{Gehrmann:1999as}
T.~Gehrmann and E.~Remiddi, \emph{{Differential equations for two loop four
  point functions}},
  \href{http://dx.doi.org/10.1016/S0550-3213(00)00223-6}{\emph{Nucl. Phys.}
  {\bf B580} (2000) 485--518}, [\href{http://arxiv.org/abs/hep-ph/9912329}{{\tt
  hep-ph/9912329}}].

\bibitem{Anastasiou:2002yz}
C.~Anastasiou and K.~Melnikov, \emph{{Higgs boson production at hadron
  colliders in NNLO QCD}},
  \href{http://dx.doi.org/10.1016/S0550-3213(02)00837-4}{\emph{Nucl. Phys.}
  {\bf B646} (2002) 220--256}, [\href{http://arxiv.org/abs/hep-ph/0207004}{{\tt
  hep-ph/0207004}}].

\bibitem{Anastasiou:2003yy}
C.~Anastasiou, L.~J. Dixon, K.~Melnikov and F.~Petriello, \emph{{Dilepton
  rapidity distribution in the Drell-Yan process at NNLO in QCD}},
  \href{http://dx.doi.org/10.1103/PhysRevLett.91.182002}{\emph{Phys. Rev.
  Lett.} {\bf 91} (2003) 182002},
  [\href{http://arxiv.org/abs/hep-ph/0306192}{{\tt hep-ph/0306192}}].

\bibitem{Anastasiou:2005cb}
C.~Anastasiou and A.~Daleo, \emph{{Numerical evaluation of loop integrals}},
  \href{http://dx.doi.org/10.1088/1126-6708/2006/10/031}{\emph{JHEP} {\bf 10}
  (2006) 031}, [\href{http://arxiv.org/abs/hep-ph/0511176}{{\tt
  hep-ph/0511176}}].

\bibitem{Ellis:2007qk}
R.~K. Ellis and G.~Zanderighi, \emph{{Scalar one-loop integrals for QCD}},
  \href{http://dx.doi.org/10.1088/1126-6708/2008/02/002}{\emph{JHEP} {\bf 02}
  (2008) 002}, [\href{http://arxiv.org/abs/0712.1851}{{\tt 0712.1851}}].

\bibitem{Smirnov:2012gma}
V.~A. Smirnov, \emph{{Analytic tools for Feynman integrals}},
  \href{http://dx.doi.org/10.1007/978-3-642-34886-0}{\emph{Springer Tracts Mod.
  Phys.} {\bf 250} (2012) 1--296}.

\bibitem{Duhr:2011zq}
C.~Duhr, H.~Gangl and J.~R. Rhodes, \emph{{From polygons and symbols to
  polylogarithmic functions}},
  \href{http://dx.doi.org/10.1007/JHEP10(2012)075}{\emph{JHEP} {\bf 10} (2012)
  075}, [\href{http://arxiv.org/abs/1110.0458}{{\tt 1110.0458}}].

\bibitem{Anastasiou:2013srw}
C.~Anastasiou, C.~Duhr, F.~Dulat and B.~Mistlberger, \emph{{Soft triple-real
  radiation for Higgs production at N3LO}},
  \href{http://dx.doi.org/10.1007/JHEP07(2013)003}{\emph{JHEP} {\bf 07} (2013)
  003}, [\href{http://arxiv.org/abs/1302.4379}{{\tt 1302.4379}}].

\bibitem{Caola:2014lpa}
F.~Caola, J.~M. Henn, K.~Melnikov and V.~A. Smirnov, \emph{{Non-planar master
  integrals for the production of two off-shell vector bosons in collisions of
  massless partons}},
  \href{http://dx.doi.org/10.1007/JHEP09(2014)043}{\emph{JHEP} {\bf 09} (2014)
  043}, [\href{http://arxiv.org/abs/1404.5590}{{\tt 1404.5590}}].

\bibitem{Henn:2014loa}
J.~M. Henn, \emph{{Multiloop integrals made simple: applications to QCD
  processes}},  in \emph{{Proceedings, 49th Rencontres de Moriond on QCD and
  High Energy Interactions: La Thuile, Italy, March 22-29, 2014}},
  pp.~289--292, 2014.
\newblock \href{http://arxiv.org/abs/1405.3683}{{\tt 1405.3683}}.

\bibitem{Henn:2014qga}
J.~M. Henn, \emph{{Lectures on differential equations for Feynman integrals}},
  \href{http://dx.doi.org/10.1088/1751-8113/48/15/153001}{\emph{J. Phys.} {\bf
  A48} (2015) 153001}, [\href{http://arxiv.org/abs/1412.2296}{{\tt
  1412.2296}}].

\bibitem{Zhang:2016kfo}
Y.~Zhang, \emph{{Lecture Notes on Multi-loop Integral Reduction and Applied
  Algebraic Geometry}},  2016.
\newblock \href{http://arxiv.org/abs/1612.02249}{{\tt 1612.02249}}.

\bibitem{Mangano:1988kk}
M.~L. Mangano, \emph{{The Color Structure of Gluon Emission}},
  \href{http://dx.doi.org/10.1016/0550-3213(88)90453-1}{\emph{Nucl. Phys.} {\bf
  B309} (1988) 461--475}.

\bibitem{Berends:1987cv}
F.~A. Berends and W.~Giele, \emph{{The Six Gluon Process as an Example of
  Weyl-Van Der Waerden Spinor Calculus}},
  \href{http://dx.doi.org/10.1016/0550-3213(87)90604-3}{\emph{Nucl. Phys.} {\bf
  B294} (1987) 700--732}.

\bibitem{Kosower:1987ic}
D.~Kosower, B.-H. Lee and V.~P. Nair, \emph{{Multi Gluon Scattering: A String
  Based Calculation}},
  \href{http://dx.doi.org/10.1016/0370-2693(88)90085-8}{\emph{Phys. Lett.} {\bf
  B201} (1988) 85--89}.

\bibitem{Mangano:1987xk}
M.~L. Mangano, S.~J. Parke and Z.~Xu, \emph{{Duality and Multi - Gluon
  Scattering}},
  \href{http://dx.doi.org/10.1016/0550-3213(88)90001-6}{\emph{Nucl. Phys.} {\bf
  B298} (1988) 653--672}.

\bibitem{Bern:1990ux}
Z.~Bern and D.~A. Kosower, \emph{{Color decomposition of one loop amplitudes in
  gauge theories}},
  \href{http://dx.doi.org/10.1016/0550-3213(91)90567-H}{\emph{Nucl. Phys.} {\bf
  B362} (1991) 389--448}.

\bibitem{Cvitanovic:1980bu}
P.~Cvitanovic, P.~G. Lauwers and P.~N. Scharbach, \emph{{Gauge Invariance
  Structure of Quantum Chromodynamics}},
  \href{http://dx.doi.org/10.1016/0550-3213(81)90098-5}{\emph{Nucl. Phys.} {\bf
  B186} (1981) 165--186}.

\bibitem{Maltoni:2002mq}
F.~Maltoni, K.~Paul, T.~Stelzer and S.~Willenbrock, \emph{{Color flow
  decomposition of QCD amplitudes}},
  \href{http://dx.doi.org/10.1103/PhysRevD.67.014026}{\emph{Phys. Rev.} {\bf
  D67} (2003) 014026}, [\href{http://arxiv.org/abs/hep-ph/0209271}{{\tt
  hep-ph/0209271}}].

\bibitem{Weinzierl:2005dd}
S.~Weinzierl, \emph{{Automated computation of spin- and colour-correlated Born
  matrix elements}},
  \href{http://dx.doi.org/10.1140/epjc/s2005-02467-6}{\emph{Eur. Phys. J.} {\bf
  C45} (2006) 745--757}, [\href{http://arxiv.org/abs/hep-ph/0510157}{{\tt
  hep-ph/0510157}}].

\bibitem{Mangano:1990by}
M.~L. Mangano and S.~J. Parke, \emph{{Multiparton amplitudes in gauge
  theories}}, \href{http://dx.doi.org/10.1016/0370-1573(91)90091-Y}{\emph{Phys.
  Rept.} {\bf 200} (1991) 301--367},
  [\href{http://arxiv.org/abs/hep-th/0509223}{{\tt hep-th/0509223}}].

\bibitem{Reuschle:2013qna}
C.~Reuschle and S.~Weinzierl, \emph{{Decomposition of one-loop QCD amplitudes
  into primitive amplitudes based on shuffle relations}},
  \href{http://dx.doi.org/10.1103/PhysRevD.88.105020}{\emph{Phys. Rev.} {\bf
  D88} (2013) 105020}, [\href{http://arxiv.org/abs/1310.0413}{{\tt
  1310.0413}}].

\bibitem{Kleiss:1988ne}
R.~Kleiss and H.~Kuijf, \emph{{Multi - Gluon Cross-sections and Five Jet
  Production at Hadron Colliders}},
  \href{http://dx.doi.org/10.1016/0550-3213(89)90574-9}{\emph{Nucl. Phys.} {\bf
  B312} (1989) 616--644}.

\bibitem{Bern:2008qj}
Z.~Bern, J.~J.~M. Carrasco and H.~Johansson, \emph{{New Relations for
  Gauge-Theory Amplitudes}},
  \href{http://dx.doi.org/10.1103/PhysRevD.78.085011}{\emph{Phys. Rev.} {\bf
  D78} (2008) 085011}, [\href{http://arxiv.org/abs/0805.3993}{{\tt
  0805.3993}}].

\bibitem{DelDuca:1999iql}
V.~Del~Duca, A.~Frizzo and F.~Maltoni, \emph{{Factorization of tree QCD
  amplitudes in the high-energy limit and in the collinear limit}},
  \href{http://dx.doi.org/10.1016/S0550-3213(99)00657-4}{\emph{Nucl. Phys.}
  {\bf B568} (2000) 211--262}, [\href{http://arxiv.org/abs/hep-ph/9909464}{{\tt
  hep-ph/9909464}}].

\bibitem{DelDuca:1999rs}
V.~Del~Duca, L.~J. Dixon and F.~Maltoni, \emph{{New color decompositions for
  gauge amplitudes at tree and loop level}},
  \href{http://dx.doi.org/10.1016/S0550-3213(99)00809-3}{\emph{Nucl. Phys.}
  {\bf B571} (2000) 51--70}, [\href{http://arxiv.org/abs/hep-ph/9910563}{{\tt
  hep-ph/9910563}}].

\bibitem{Johansson:2015oia}
H.~Johansson and A.~Ochirov, \emph{{Color-Kinematics Duality for QCD
  Amplitudes}}, \href{http://dx.doi.org/10.1007/JHEP01(2016)170}{\emph{JHEP}
  {\bf 01} (2016) 170}, [\href{http://arxiv.org/abs/1507.00332}{{\tt
  1507.00332}}].

\bibitem{Melia:2015ika}
T.~Melia, \emph{{Proof of a new colour decomposition for QCD amplitudes}},
  \href{http://dx.doi.org/10.1007/JHEP12(2015)107}{\emph{JHEP} {\bf 12} (2015)
  107}, [\href{http://arxiv.org/abs/1509.03297}{{\tt 1509.03297}}].

\bibitem{Bern:1994fz}
Z.~Bern, L.~J. Dixon and D.~A. Kosower, \emph{{One loop corrections to two
  quark three gluon amplitudes}},
  \href{http://dx.doi.org/10.1016/0550-3213(94)00542-M}{\emph{Nucl. Phys.} {\bf
  B437} (1995) 259--304}, [\href{http://arxiv.org/abs/hep-ph/9409393}{{\tt
  hep-ph/9409393}}].

\bibitem{Ita:2011ar}
H.~Ita and K.~Ozeren, \emph{{Colour Decompositions of Multi-quark One-loop QCD
  Amplitudes}}, \href{http://dx.doi.org/10.1007/JHEP02(2012)118}{\emph{JHEP}
  {\bf 02} (2012) 118}, [\href{http://arxiv.org/abs/1111.4193}{{\tt
  1111.4193}}].

\bibitem{Badger:2012pg}
S.~Badger, B.~Biedermann, P.~Uwer and V.~Yundin, \emph{{Numerical evaluation of
  virtual corrections to multi-jet production in massless QCD}},
  \href{http://dx.doi.org/10.1016/j.cpc.2013.03.018}{\emph{Comput. Phys.
  Commun.} {\bf 184} (2013) 1981--1998},
  [\href{http://arxiv.org/abs/1209.0100}{{\tt 1209.0100}}].

\bibitem{Schuster:2013aya}
T.~Schuster, \emph{{Color ordering in QCD}},
  \href{http://dx.doi.org/10.1103/PhysRevD.89.105022}{\emph{Phys. Rev.} {\bf
  D89} (2014) 105022}, [\href{http://arxiv.org/abs/1311.6296}{{\tt
  1311.6296}}].

\bibitem{Ochirov:2016ewn}
A.~Ochirov and B.~Page, \emph{{Full Colour for Loop Amplitudes in Yang-Mills
  Theory}}, \href{http://dx.doi.org/10.1007/JHEP02(2017)100}{\emph{JHEP} {\bf
  02} (2017) 100}, [\href{http://arxiv.org/abs/1612.04366}{{\tt 1612.04366}}].

\bibitem{Dixon:1996wi}
L.~J. Dixon, \emph{{Calculating scattering amplitudes efficiently}},  in
  \emph{{QCD and beyond. Proceedings, Theoretical Advanced Study Institute in
  Elementary Particle Physics, TASI-95, Boulder, USA, June 4-30, 1995}},
  pp.~539--584, 1996.
\newblock \href{http://arxiv.org/abs/hep-ph/9601359}{{\tt hep-ph/9601359}}.

\bibitem{Melia:2013bta}
T.~Melia, \emph{{Dyck words and multiquark primitive amplitudes}},
  \href{http://dx.doi.org/10.1103/PhysRevD.88.014020}{\emph{Phys. Rev.} {\bf
  D88} (2013) 014020}, [\href{http://arxiv.org/abs/1304.7809}{{\tt
  1304.7809}}].

\bibitem{Melia:2014oza}
T.~Melia, \emph{{Dyck words and multi-quark amplitudes}}, {\emph{PoS} {\bf
  RADCOR2013} (2013) 031}.

\bibitem{Duchon:2000}
P.~Duchon, \emph{On the enumeration and generation of generalized dyck words},
  \href{http://dx.doi.org/https://doi.org/10.1016/S0012-365X(00)00150-3}{\emph{Discrete
  Mathematics} {\bf 225} (2000) 121 -- 135}.

\bibitem{Kasa:2010}
Z.~K{\'{a}}sa, \emph{Generating and ranking of dyck words}, {\emph{CoRR} {\bf
  abs/1002.2625} (2010) }, [\href{http://arxiv.org/abs/1002.2625}{{\tt
  1002.2625}}].

\bibitem{Melia:2013epa}
T.~Melia, \emph{{Getting more flavor out of one-flavor QCD}},
  \href{http://dx.doi.org/10.1103/PhysRevD.89.074012}{\emph{Phys. Rev.} {\bf
  D89} (2014) 074012}, [\href{http://arxiv.org/abs/1312.0599}{{\tt
  1312.0599}}].

\bibitem{mario}
``{Mario Bros}.'' [Nintendo Arcade], 1983.

\bibitem{Carrasco:2011hw}
J.~J.~M. Carrasco and H.~Johansson, \emph{{Generic multiloop methods and
  application to N=4 super-Yang-Mills}},
  \href{http://dx.doi.org/10.1088/1751-8113/44/45/454004}{\emph{J. Phys.} {\bf
  A44} (2011) 454004}, [\href{http://arxiv.org/abs/1103.3298}{{\tt
  1103.3298}}].

\bibitem{Bern:2011qt}
Z.~Bern and Y.-t. Huang, \emph{{Basics of Generalized Unitarity}},
  \href{http://dx.doi.org/10.1088/1751-8113/44/45/454003}{\emph{J. Phys.} {\bf
  A44} (2011) 454003}, [\href{http://arxiv.org/abs/1103.1869}{{\tt
  1103.1869}}].

\bibitem{Ita:2011hi}
H.~Ita, \emph{{Susy Theories and QCD: Numerical Approaches}},
  \href{http://dx.doi.org/10.1088/1751-8113/44/45/454005}{\emph{J. Phys.} {\bf
  A44} (2011) 454005}, [\href{http://arxiv.org/abs/1109.6527}{{\tt
  1109.6527}}].

\bibitem{Britto:2010xq}
R.~Britto, \emph{{Loop Amplitudes in Gauge Theories: Modern Analytic
  Approaches}},
  \href{http://dx.doi.org/10.1088/1751-8113/44/45/454006}{\emph{J. Phys.} {\bf
  A44} (2011) 454006}, [\href{http://arxiv.org/abs/1012.4493}{{\tt
  1012.4493}}].

\bibitem{NigelGlover:2008ur}
E.~W. Nigel~Glover and C.~Williams, \emph{{One-Loop Gluonic Amplitudes from
  Single Unitarity Cuts}},
  \href{http://dx.doi.org/10.1088/1126-6708/2008/12/067}{\emph{JHEP} {\bf 12}
  (2008) 067}, [\href{http://arxiv.org/abs/0810.2964}{{\tt 0810.2964}}].

\bibitem{Bierenbaum:2010cy}
I.~Bierenbaum, S.~Catani, P.~Draggiotis and G.~Rodrigo, \emph{{A Tree-Loop
  Duality Relation at Two Loops and Beyond}},
  \href{http://dx.doi.org/10.1007/JHEP10(2010)073}{\emph{JHEP} {\bf 10} (2010)
  073}, [\href{http://arxiv.org/abs/1007.0194}{{\tt 1007.0194}}].

\bibitem{Elvang:2011ub}
H.~Elvang, D.~Z. Freedman and M.~Kiermaier, \emph{{Integrands for QCD rational
  terms and N=4 SYM from massive CSW rules}},
  \href{http://dx.doi.org/10.1007/JHEP06(2012)015}{\emph{JHEP} {\bf 06} (2012)
  015}, [\href{http://arxiv.org/abs/1111.0635}{{\tt 1111.0635}}].

\bibitem{CaronHuot:2010zt}
S.~Caron-Huot, \emph{{Loops and trees}},
  \href{http://dx.doi.org/10.1007/JHEP05(2011)080}{\emph{JHEP} {\bf 05} (2011)
  080}, [\href{http://arxiv.org/abs/1007.3224}{{\tt 1007.3224}}].

\bibitem{Boels:2010nw}
R.~H. Boels, \emph{{On BCFW shifts of integrands and integrals}},
  \href{http://dx.doi.org/10.1007/JHEP11(2010)113}{\emph{JHEP} {\bf 11} (2010)
  113}, [\href{http://arxiv.org/abs/1008.3101}{{\tt 1008.3101}}].

\bibitem{Baadsgaard:2015twa}
C.~Baadsgaard, N.~E.~J. Bjerrum-Bohr, J.~L. Bourjaily, S.~Caron-Huot, P.~H.
  Damgaard and B.~Feng, \emph{{New Representations of the Perturbative
  S-Matrix}},
  \href{http://dx.doi.org/10.1103/PhysRevLett.116.061601}{\emph{Phys. Rev.
  Lett.} {\bf 116} (2016) 061601}, [\href{http://arxiv.org/abs/1509.02169}{{\tt
  1509.02169}}].

\bibitem{Huang:2015cwh}
R.~Huang, Q.~Jin, J.~Rao, K.~Zhou and B.~Feng, \emph{{The Q-cut Representation
  of One-loop Integrands and Unitarity Cut Method}},
  \href{http://dx.doi.org/10.1007/JHEP03(2016)057}{\emph{JHEP} {\bf 03} (2016)
  057}, [\href{http://arxiv.org/abs/1512.02860}{{\tt 1512.02860}}].

\bibitem{AlvarezGaume:1983ig}
L.~Alvarez-Gaume and E.~Witten, \emph{{Gravitational Anomalies}},
  \href{http://dx.doi.org/10.1016/0550-3213(84)90066-X}{\emph{Nucl. Phys.} {\bf
  B234} (1984) 269}.

\bibitem{ArkaniHamed:2010gh}
N.~Arkani-Hamed, J.~L. Bourjaily, F.~Cachazo and J.~Trnka, \emph{{Local
  Integrals for Planar Scattering Amplitudes}},
  \href{http://dx.doi.org/10.1007/JHEP06(2012)125}{\emph{JHEP} {\bf 06} (2012)
  125}, [\href{http://arxiv.org/abs/1012.6032}{{\tt 1012.6032}}].

\bibitem{Drummond:2006rz}
J.~M. Drummond, J.~Henn, V.~A. Smirnov and E.~Sokatchev, \emph{{Magic
  identities for conformal four-point integrals}},
  \href{http://dx.doi.org/10.1088/1126-6708/2007/01/064}{\emph{JHEP} {\bf 01}
  (2007) 064}, [\href{http://arxiv.org/abs/hep-th/0607160}{{\tt
  hep-th/0607160}}].

\bibitem{Bern:2010ue}
Z.~Bern, J.~J.~M. Carrasco and H.~Johansson, \emph{{Perturbative Quantum
  Gravity as a Double Copy of Gauge Theory}},
  \href{http://dx.doi.org/10.1103/PhysRevLett.105.061602}{\emph{Phys. Rev.
  Lett.} {\bf 105} (2010) 061602}, [\href{http://arxiv.org/abs/1004.0476}{{\tt
  1004.0476}}].

\bibitem{Brown:2016hck}
R.~W. Brown and S.~G. Naculich, \emph{{Color-factor symmetry and BCJ relations
  for QCD amplitudes}},
  \href{http://dx.doi.org/10.1007/JHEP11(2016)060}{\emph{JHEP} {\bf 11} (2016)
  060}, [\href{http://arxiv.org/abs/1608.05291}{{\tt 1608.05291}}].

\bibitem{Brown:2016mrh}
R.~W. Brown and S.~G. Naculich, \emph{{BCJ relations from a new symmetry of
  gauge-theory amplitudes}},
  \href{http://dx.doi.org/10.1007/JHEP10(2016)130}{\emph{JHEP} {\bf 10} (2016)
  130}, [\href{http://arxiv.org/abs/1608.04387}{{\tt 1608.04387}}].

\bibitem{Boels:2013bi}
R.~H. Boels, R.~S. Isermann, R.~Monteiro and D.~O'Connell,
  \emph{{Colour-Kinematics Duality for One-Loop Rational Amplitudes}},
  \href{http://dx.doi.org/10.1007/JHEP04(2013)107}{\emph{JHEP} {\bf 04} (2013)
  107}, [\href{http://arxiv.org/abs/1301.4165}{{\tt 1301.4165}}].

\bibitem{He:2015wgf}
S.~He, R.~Monteiro and O.~Schlotterer, \emph{{String-inspired BCJ numerators
  for one-loop MHV amplitudes}},
  \href{http://dx.doi.org/10.1007/JHEP01(2016)171}{\emph{JHEP} {\bf 01} (2016)
  171}, [\href{http://arxiv.org/abs/1507.06288}{{\tt 1507.06288}}].

\bibitem{Primo:2016omk}
A.~Primo and W.~J. Torres~Bobadilla, \emph{{BCJ Identities and $d$-Dimensional
  Generalized Unitarity}},
  \href{http://dx.doi.org/10.1007/JHEP04(2016)125}{\emph{JHEP} {\bf 04} (2016)
  125}, [\href{http://arxiv.org/abs/1602.03161}{{\tt 1602.03161}}].

\bibitem{Tourkine:2016bak}
P.~Tourkine and P.~Vanhove, \emph{{Higher-loop amplitude monodromy relations in
  string and gauge theory}},
  \href{http://dx.doi.org/10.1103/PhysRevLett.117.211601}{\emph{Phys. Rev.
  Lett.} {\bf 117} (2016) 211601}, [\href{http://arxiv.org/abs/1608.01665}{{\tt
  1608.01665}}].

\bibitem{He:2016mzd}
S.~He and O.~Schlotterer, \emph{{New Relations for Gauge-Theory and Gravity
  Amplitudes at Loop Level}},
  \href{http://dx.doi.org/10.1103/PhysRevLett.118.161601}{\emph{Phys. Rev.
  Lett.} {\bf 118} (2017) 161601}, [\href{http://arxiv.org/abs/1612.00417}{{\tt
  1612.00417}}].

\bibitem{He:2017spx}
S.~He, O.~Schlotterer and Y.~Zhang, \emph{{New BCJ representations for one-loop
  amplitudes in gauge theories and gravity}},
  \href{http://arxiv.org/abs/1706.00640}{{\tt 1706.00640}}.

\bibitem{Hohenegger:2017kqy}
S.~Hohenegger and S.~Stieberger, \emph{{Monodromy Relations in Higher-Loop
  String Amplitudes}},
  \href{http://dx.doi.org/10.1016/j.nuclphysb.2017.09.020}{\emph{Nucl. Phys.}
  {\bf B925} (2017) 63--134}, [\href{http://arxiv.org/abs/1702.04963}{{\tt
  1702.04963}}].

\bibitem{Ochirov:2017jby}
A.~Ochirov, P.~Tourkine and P.~Vanhove, \emph{{One-loop monodromy relations on
  single cuts}}, \href{http://dx.doi.org/10.1007/JHEP10(2017)105}{\emph{JHEP}
  {\bf 10} (2017) 105}, [\href{http://arxiv.org/abs/1707.05775}{{\tt
  1707.05775}}].

\bibitem{Mafra:2017ioj}
C.~R. Mafra and O.~Schlotterer, \emph{{The double-copy structure of one-loop
  open-string amplitudes}},  \href{http://arxiv.org/abs/1711.09104}{{\tt
  1711.09104}}.

\bibitem{Geyer:2017ela}
Y.~Geyer and R.~Monteiro, \emph{{Gluons and gravitons at one loop from
  ambitwistor strings}},  \href{http://arxiv.org/abs/1711.09923}{{\tt
  1711.09923}}.

\bibitem{Jurado:2017xut}
J.~L. Jurado, G.~Rodrigo and W.~J. Torres~Bobadilla, \emph{{From Jacobi
  off-shell currents to integral relations}},
  \href{http://dx.doi.org/10.1007/JHEP12(2017)122}{\emph{JHEP} {\bf 12} (2017)
  122}, [\href{http://arxiv.org/abs/1710.11010}{{\tt 1710.11010}}].

\bibitem{Johansson:2017bfl}
H.~Johansson, G.~Kälin and G.~Mogull, \emph{{Two-loop supersymmetric QCD and
  half-maximal supergravity amplitudes}},
  \href{http://dx.doi.org/10.1007/JHEP09(2017)019}{\emph{JHEP} {\bf 09} (2017)
  019}, [\href{http://arxiv.org/abs/1706.09381}{{\tt 1706.09381}}].

\end{thebibliography}\endgroup
